# From stacking to function: emergent states and quantum devices in 2D superconductor heterostructures

Sichun Zhao(赵思莼)[1,†], Junlin Xiong(熊俊林)[1,†], Ji Zhou(周吉)[1], Shi-Jun Liang(梁世军)[1,2], Bin Cheng(程斌)[1,2,3,*], Feng Miao(缪峰)[1,2,*]

[1]Institute of Brain-Inspired Intelligence, National Laboratory of Solid State Microstructures, School of Physics, Collaborative Innovation Center of Advanced Microstructures, Nanjing University, Nanjing 210093, China.

[2]Jiangsu Physical Science Research Center, Nanjing 210093, China.

[3]Institute of Interdisciplinary Physical Sciences, School of Science, Nanjing University of Science and Technology, Nanjing 210094, China.

\* Corresponding author. Email: bincheng@njust.edu.cn; miao@nju.edu.cn

† These authors contributed equally to this work.

**Abstract:** Two-dimensional (2D) superconductors provide a powerful building block for engineering emergent quantum states shaped by reduced dimensionality, enhanced quantum fluctuations, and interfacial symmetry breaking. In van der Waals (vdW) heterostructures, atomically sharp and lattice-mismatch-free interfaces enable superconductivity to be deliberately coupled with magnetism, spin–orbit interaction, and band topology, allowing collective electronic orders to be combined and reconfigured in ways unattainable in bulk materials. This Review summarizes recent advances in vdW heterostructures of 2D superconductors, focusing on superconductor/magnet (S/M), superconductor/topological material (S/T), and superconductor/superconductor (S/S) junctions. We discuss the microscopic mechanisms underlying proximity effects and highlight how interfacial exchange fields, spin–orbit coupling, and twist-controlled tunneling give rise to unconventional pairing, long-range spin-triplet supercurrents, nonreciprocal Josephson transport, and topological superconductivity potentially hosting Majorana bound states. Beyond their fundamental significance, the ability to controllably generate topological and nonreciprocal superconducting states positions 2D superconductor heterostructures as promising building blocks for emerging quantum technologies, including ultra-sensitive quantum sensing, programmable superconducting logic, and energy-efficient quantum and neuromorphic computing architectures. Looking forward, advances in materials synthesis, interface engineering, and device integration are expected to further expand the scope and functionality of 2D superconductor heterostructures, reinforcing their role as a central platform for exploring and controlling emergent quantum phases.

## 1. Introduction

Two-dimensional (2D) superconductors provide an exceptionally versatile platform for exploring unconventional quantum states in which reduced dimensionality,[1-4] enhanced quantum fluctuations,[5] and interfacial effects[6-9] fundamentally reshape the nature of electronic order. In such systems, superconductivity becomes intrinsically intertwined with spin polarization,[10] spin–orbit coupling (SOC),[11,12] and topological band geometry[13,14] at the atomic scale. These features make 2D materials an ideal building block for engineering superconductor heterostructures in which distinct orders—superconductivity, magnetism, and topology—can be deliberately combined to generate emergent quantum phenomena unattainable in conventional three-dimensional compounds.

Superconductivity, ferromagnetism, and topological order represent three distinct forms of collective electronic behavior, each characterized by its own order parameter and underlying symmetry. Their coexistence and mutual coupling provide a powerful route toward emergent quantum states. Enabled by broken symmetries,[15] interfacial SOC,[16-18] and quantum coherence, the interplay among these orders can be either competitive or cooperative, giving rise to unconventional pairing symmetries,[19-23] and exotic boundary excitations[24-29] such as Majorana zero modes (MZMs). A paradigmatic example is the antagonism between conventional spin-singlet superconductivity and ferromagnetism:[30,31] the exchange field of a ferromagnet tends to destroy singlet Cooper pairs, preventing their homogeneous coexistence. However, when superconductivity and magnetism meet at an interface with strong SOC and broken inversion symmetry, singlet pairs can be converted into equal-spin triplet correlations.[32-38] These triplet pairs penetrate ferromagnets over long distances and support long-range dissipationless supercurrents,[39-41] providing a viable pathway toward superconducting spintronics. Beyond transport, such triplet-enriched proximity effects can drive hybrid structures into topological regimes that host Majorana modes[42-44]—non-Abelian boundary excitations with intrinsic resilience against local perturbations and direct relevance to topological quantum computation. Closely related physics emerges at interfaces between superconductors and topological materials. The Fu–Kane mechanism[45] established that coupling a conventional s-wave superconductor to a topological insulator can induce a topological superconducting gap supporting Majorana bound states. More broadly, S/T heterostructures formed with topological insulators or topological semimetals enable a rich spectrum of unconventional superconducting states, including topological hinge[46] and edge superconductivity, and topologically protected boundary modes. These systems offer a controllable route to engineering topological superconductivity without relying on intrinsically complex correlated materials.

The interplay among superconductivity, magnetism, and topology provides a powerful route toward emergent quantum phenomena arising from the coupling of distinct collective orders. Van der Waals (vdW) superconductor heterostructures offer a particularly versatile and controllable platform for exploring such coupled states. Rather than relying on the intrinsic coexistence of multiple orders within a single compound, vdW heterostructures enable superconductivity, magnetism, and topological band structures to be deliberately assembled and interfaced through proximity effects across atomically sharp junctions. A key advantage of vdW heterostructures lies in their atomically flat, dangling-bond-free interfaces, which naturally ensure high interfacial quality without the need for lattice matching or complex epitaxial growth.[47-51] This structural simplicity enhances electronic transparency and coherence across interfaces, both of which are essential for efficient proximity coupling. At the same time, the 2D geometry of vdW materials allows additional external control through electrostatic gating,[52-61] interlayer twist-angle engineering,[62-67] and strain,[68-71] providing an exceptional level of tunability over superconducting order parameters, spin textures, and topological band structures. Together, these features make vdW superconductor heterostructures a uniquely powerful platform for realizing unconventional pairing states, long-range Josephson coupling, nonreciprocal superconducting transport, and topological boundary modes.

In this review, we summarize recent progress on emergent phenomena in 2D superconductor heterostructures, focusing on three representative classes: superconductor/magnet (S/M), superconductor/topological material (S/T), and superconductor/superconductor (S/S) junctions (Fig.1). We highlight the underlying interfacial coupling mechanisms that govern their collective behavior, summarize recent experimental breakthroughs, and discuss their implications for spin-triplet pairing, chiral superconductivity, unconventional Josephson effects, and Majorana-based quantum devices. Beyond their fundamental significance, these advances also point toward a broad range of future applications,[72-75] including programmable superconducting logic elements, quantum computing circuits, in-memory and neuromorphic quantum computing architectures, and quantum intelligent sensing. This review aims to provide a conceptual and methodological framework for future studies of 2D superconductivity and its heterostructure-enabled quantum phases, bridging fundamental condensed-matter physics and emerging quantum device technologies.

# 2. Two-dimensional superconducting materials as intrinsic and engineered building blocks for heterostructures

Two-dimensional superconducting materials form the essential building blocks for constructing superconductor heterostructures. As the system dimensionality is reduced from three to two dimensions, quantum fluctuations are strongly enhanced, dielectric screening is suppressed, and electronic states become highly sensitive to interfaces and external perturbations. These effects profoundly reshape Cooper pairing, collective modes, and competing electronic orders. In the two-dimensional limit, superconductivity is exceptionally tunable via electrostatic gating, strain engineering, twist-angle control, and proximity coupling, enabling precise manipulation of the superconducting phase, spin texture, and band topology. Together, these features provide a versatile and programmable materials basis for designing heterostructures with tailored quantum functionalities.

## 2.1. Transition metal dichalcogenides

Layered transition-metal dichalcogenides (TMDCs), such as $NbSe_2$ and $TaS_2$, represent one of the most extensively studied classes of intrinsic 2D superconductors. Their crystal structures consist of strongly bonded in-plane atomic layers and weakly coupled vdW interlayers,[76] enabling mechanical exfoliation down to the monolayer limit while preserving high crystalline quality. As a result, TMDCs provide an ideal materials platform for investigating how superconductivity evolves under reduced dimensionality and enhanced interfacial sensitivity.

Among them, $NbSe_2$ stands out as a paradigmatic system. Remarkably, its superconductivity remains robust down to a single atomic layer ($T_c$ (N=1) ≈ 3 K), while simultaneously giving rise to a variety of emergent phenomena that are absent in the bulk. For example, the combination of strong intrinsic SOC and broken inversion symmetry in monolayer $NbSe_2$ generates an Ising-type spin–orbit interaction that locks electron spins perpendicular to the atomic plane[77] (Fig. 2(a)). This spin locking strongly suppresses in-plane spin polarization of Cooper pairs, thereby stabilizing superconductivity against large in-plane magnetic fields. As a consequence, the in-plane upper critical field dramatically exceeds the conventional Pauli paramagnetic limit (Fig. 2(b)). Dimensional reduction in TMDCs also profoundly affects competing electronic orders. In $NbSe_2$, charge-density-wave (CDW) order is significantly enhanced as the thickness is reduced, with spectroscopic measurements revealing a substantial increase in the CDW transition temperature accompanied by the opening of a sizable energy gap[78] (Fig. 2(c)). The coexistence and competition between superconductivity and CDW order highlight the delicate balance of collective electronic instabilities in the two-dimensional limit and underscore the sensitivity of superconductivity to lattice, electronic, and symmetry constraints. Beyond conventional s-wave pairing, monolayer and few-layer TMDCs have been theoretically predicted and experimentally suggested to host unconventional superconducting states enabled by strong SOC and broken inversion symmetry. These include mixed singlet–triplet pairing channels,[79] finite-momentum Cooper pairing reminiscent of Fulde–Ferrell–Larkin–Ovchinnikov (FFLO) states[80] (Fig. 2(d)), nematic superconductivity,[81] and possible topological superconducting signatures.[82-84] Together, these observations position TMDCs as a benchmark platform for exploring symmetry-broken and unconventional superconductivity in two dimensions.

Two-dimensional superconductivity can be accessed and tuned using a variety of external and engineering approaches.[85-87] Among these, electrostatic gating—particularly electric-double-layer and ionic-liquid gating—has emerged as an exceptionally powerful approach, as it enables the accumulation of extremely high carrier densities at surfaces or interfaces, thereby inducing superconductivity in otherwise semiconducting or insulating materials. A prototypical example is ionic-liquid-gated layered ZrNCl, in which superconductivity with a transition temperature of approximately 15.2 K can be electrostatically induced and a controlled crossover from BCS-like to BEC-like pairing regimes by tuning carrier density.[52,88] Similar interfacial superconductivity has also been realized in ionic-liquid-gated 1T–$SnSe_2$, where interfacial superconductivity has been realized with transition temperatures of several kelvin[60] (Fig. 2(e)). The induced superconducting phase exhibits in-plane critical fields that exceed the Pauli limit, suggesting a gate-tunable interplay between carrier density, SOC, and superconducting pairing. $MoS_2$ constitute another model system,[56,57,89] in which robust two-dimensional superconductivity emerges under electric-double-layer gating. Specifically, this superconducting state can be experimentally observed at relatively moderate carrier densities (Fig. 2(f)). First-principles calculations indicate that the softening of acoustic phonon modes near the Brillouin-zone M point plays a crucial role in enhancing electron–phonon coupling and stabilizing the superconducting state.

Beyond electrostatic control, dimensionality itself provides a powerful and conceptually distinct tuning parameter for accessing unconventional superconducting phases. Recent work has reported dimensionality-tunable zero-field anomalous metallic states (AMS) observed under carefully implemented radio-frequency filtering near superconducting quantum phase transitions in the layered TMDC superconductor 2H–$Ta_2S_3Se$.[90] The AMS is characterized by a resistance that saturates to a finite, temperature-independent plateau at low temperatures.[54,91-95] Early transport studies[96] on crystalline 2H–$NbSe_2$ suggested that such apparent resistance saturation could be eliminated by rigorous filtering, implying that two-dimensional superconductivity may be extremely sensitive to extrinsic perturbations. However, subsequent experiments demonstrated that even after suppressing high-frequency external noise, AMS can persist at low temperatures under strong magnetic fields. Besides, measurements on 4Ha–$TaSe_2$ nanodevices,[97] combining rigorous radiation filtering with ultralow-temperature measurements, provided compelling evidence for an intrinsic AMS, indicating that a genuine metallic ground state may exist in crystalline two-dimensional superconductors under appropriate conditions. Within this context, the AMS observed in 2H–$Ta_2S_3Se$ is particularly notable, as it is measured under stringent radio-frequency filtering conditions and thus cannot be attributed to uncontrolled external radiation. As the thickness is reduced below 10 nm, a magnetic-field-driven AMS emerges, characterized by a vanishing Hall resistance coexisting with a finite longitudinal resistance. Remarkably, upon further reducing the thickness to the few-layer limit (~3 nm), an unexpected zero-field AMS appears (Fig. 2(g)). This state is consistent with a quantum vortex creep picture and exhibits pronounced nonreciprocal transport behavior, suggesting the onset of spontaneous time-reversal symmetry breaking accompanied by vortex dynamics as the system approaches the two-dimensional limit. More broadly, a growing set of control knobs—including interfacial charge transfer, strain engineering, and moiré superlattice design in twisted multilayer systems—has enabled superconductivity and emergent transport phenomena in an expanding family of artificial two-dimensional platforms. Collectively, these developments demonstrate that superconductivity in TMDC-based systems can be treated as a programmable quantum state, offering an unprecedented materials playground for heterostructure engineering and for exploring emergent superconducting phenomena in reduced dimensions.

## 2.2. High-temperature cuprate superconductors

High-temperature cuprate superconductors constitute a central class of quasi-two-dimensional superconductors, whose essential electronic structure is governed by $CuO_2$ planes. Superconductivity emerges upon carrier doping of a Mott-insulating parent compound, producing a characteristic phase diagram in which a superconducting phase intertwines with antiferromagnetism, CDW order, and the pseudogap regime.[98-106] These materials are paradigms of strong electron correlations, in which the pairing mechanism departs fundamentally from the conventional phonon-mediated Bardeen–Cooper–Schrieffer (BCS) framework. The superconducting order parameter in cuprates exhibits a well-established *d*-wave symmetry, giving rise to nodes in the superconducting gap and highly anisotropic quasiparticle dynamics. Together with their exceptionally high transition temperatures and extreme electronic anisotropy, cuprates provide a uniquely powerful platform for investigating unconventional pairing, competing orders, and quantum criticality in reduced dimensions.

From a materials perspective, the cuprate family encompasses several canonical systems, including La-based compounds such as $La_{2-x}M_xCuO_4$[107] (M = Sr, Ba), Y-based cuprates such as $YBa_2Cu_3O_{7-x}$[108] (YBCO), and Bi-based cuprates of the form $Bi_2Sr_2Ca_{n-1}Cu_nO_{2n+4+x}$.[109] Among them, $Bi_2Sr_2CaCu_2O_{8+x}$[110] (BSCCO) (Fig. 3(a)) occupies a particularly prominent position in the context of two-dimensional superconductivity. Owing to its extremely weak interlayer coupling and natural cleavage planes, BSCCO can be mechanically exfoliated down to the atomic limit, enabling direct access to intrinsic two-dimensional high-$T_c$ superconductivity.[110-115] Recent experimental advances have successfully realized atomically thin BSCCO flakes ($T_c \approx 85$ K) as genuine two-dimensional superconductors[110] (Fig. 3(b)). These systems exhibit a range of emergent phenomena that are inaccessible in bulk cuprates, including superconducting diode effects[116] (Fig. 3(c)) and doping-tunable quantum transition[117,118] (Fig. 3(d)). The reduced dimensionality, combined with strong correlations and broken symmetries at surfaces and interfaces, provides new routes to control superconducting phenomena in cuprate-based systems, for example through electric-field tuning of cuprates themselves[115] and through proximity-induced superconductivity in adjacent materials interfaced with cuprate superconductors.[119,120] It should be noted, however, that the nature of the reported proximity-induced superconductivity in such heterostructures remains under debate. In particular, the limited momentum and spin resolution of transport and scanning tunneling microscopy measurements makes it challenging to unambiguously establish superconductivity in topological surface states, as the observed signatures may also originate from bulk states, trivial surface bands, or impurity-induced states.[121] Taken together, two-dimensional cuprate superconductors—exemplified by exfoliated BSCCO—establish a crucial bridge between strongly correlated electron systems and two-dimensional superconducting electronics. They not only provide a fertile testing ground for long-standing questions regarding unconventional pairing and competing orders, but also offer unique opportunities for integrating high-$T_c$ superconductivity into engineered heterostructures, Josephson devices, and nonreciprocal superconducting circuits discussed in subsequent sections.

## 2.3. Iron-based superconductors

Magnetism and superconductivity were long regarded as fundamentally incompatible order parameters, as magnetic moments disrupt the spin-antiparallel configuration of Cooper pairs, rendering materials containing magnetic transition-metal ions generally unfavorable for superconductivity. This conventional view was overturned by the discovery of iron-based superconductors, in which superconductivity is well preserved despite the presence of magnetic iron ions, with transition temperatures exceeding 50 K and even reaching 60–70 K in certain interface-engineered systems. This breakthrough reshaped the understanding of the interplay between magnetism and superconductivity and opened new avenues for exploring unconventional pairing mechanisms. Since the initial discovery of superconductivity at 26 K in LaFeAsO$_{1-x}$F$_x$[122] (Fig. 4(a)) in 2008, a diverse family of FeSCs has rapidly emerged. Rare-earth substituted 1111 compounds, such as SmFeAsO$_{1-x}$F$_x$[123] and NdFeAsO$_{1-x}$F$_x$,[124] soon achieved transition temperatures in the 40–55 K range. This was followed by the discovery of the 122 family[125] (AFe$_2$As$_2$, A = Ba, Sr, Ca, K), the 111 family (LiFeAs[126,127] (Fig. 4(b)) and NaFeAs[128]), and the structurally simplest 11 family (Fig. 4(c)), consisting of FeSe and Fe(Se,Te).[129] Together, these systems established iron-based superconductivity as one of the most versatile and tunable platforms for exploring correlated quantum matter.

A defining characteristic of FeSCs is their multi-orbital, multi-band electronic structure[130] (Fig. 4(d)). Their Fermi surfaces typically comprise both hole and electron pockets, accompanied by strong orbital-dependent correlations. The parent compounds host a rich hierarchy of intertwined electronic orders,[131,132] including nematicity, spin-density-wave order, double-stripe magnetism, C$_4$-symmetric magnetic phases, and superconductivity[133] (see the phase diagram of a typical FeSC BaFe$_2$As$_2$ in Fig. 4(e)). The close proximity and competition among these collective instabilities generate strong quantum fluctuations, which are widely believed to play a central role in mediating unconventional superconducting pairing. The role of dimensionality in modulating superconducting properties is also an intriguing theme in understanding the behavior of iron-based superconductors.[134] From the perspective of reduced dimensionality and interface physics, FeSCs have demonstrated particularly striking behavior. A landmark example is monolayer FeSe grown by molecular beam epitaxy on SrTiO$_3$ substrates. In this system, scanning tunneling spectroscopy (STS) reveals a superconducting gap as large as ~22 meV,[7] indicating a dramatically enhanced pairing scale compared with bulk FeSe. Angle-resolved photoemission spectroscopy (ARPES) measurements further suggest superconducting or superconductivity-related gap features persisting up to 65–70 K.[135,136] These observations highlight the FeSe/SrTiO$_3$ interface as a paradigmatic example of interface-enhanced superconductivity, where electron–phonon coupling, charge transfer, and interfacial screening collectively reshape the superconducting state.

Among the FeSC family, iron chalcogenides are especially notable for their quasi-two-dimensional Ch–Fe–Ch (Ch = chalcogen) layered structure and their propensity to host coexisting superconductivity, magnetism, and nontrivial band topology. In particular, substituting Te into FeSe reduces the intralayer hopping while enhancing the interlayer hopping in Fe(Se,Te), leading to a band inversion and a corresponding change in the system's topological character.[137] With further increase of the Te concentration, enhanced antiferromagnetic spin fluctuations emerge due to the proximity to the FeTe magnetic phase,[138] establishing Fe(Se,Te)[130] (Fig. 4(c)) as an archetypal platform for intrinsic topological superconductivity. The observation of Dirac-cone–like surface states in ARPES measurements (Fig. 4(f)) provides possible evidence for the presence of topological surface states. Complementary probes, including nitrogen-vacancy center magnetometry,[139] polar Kerr effect measurements,[140] and muon spin rotation (µSR) experiments,[138] further reveal the coexistence of local magnetism and superconductivity (Fig. 4(g)). Moreover, scanning tunneling microscopy studies report robust zero-bias conductance peaks at vortex cores that are consistent with Majorana bound states[141] (Fig. 4(h)). In addition to these vortex-core states, zero-energy bound states have also been observed at both ends of one-dimensional atomic line defects in FeTe$_{0.5}$Se$_{0.5}$ films grown on SrTiO$_3$(001) substrates,[142] providing an alternative platform for realizing Majorana bound states. Collectively, iron-based superconductors—spanning intrinsic bulk materials, atomically thin films, and interface-engineered heterostructures—occupy a unique position at the intersection of magnetism, topology, and superconductivity. Their multi-orbital nature, strong correlations, and sensitivity to dimensionality render them an essential materials foundation for superconductor heterostructures, in which superconducting order can be interfaced, hybridized, and topologically reconfigured through proximity to magnetic, topological, and spin–orbit-coupled materials, enabling emergent quantum states unattainable in isolated components.

## 3. Superconductor/magnet (S/M) heterostructures

Two-dimensional superconductor/magnet (S/M) heterostructures provide an exceptionally rich platform for investigating the mutual antagonism and cooperation between superconductivity and magnetism at the ultimate thickness limit. In conventional three-dimensional systems, these two electronic orders are largely incompatible because the exchange field of a ferromagnet breaks time-reversal symmetry and suppresses spin-singlet Cooper pairing. In contrast, reduced dimensionality and atomically sharp interfaces in 2D materials dramatically enhance proximity effects and interfacial SOC, enabling new regimes of coexistence and emergent superconducting states that are inaccessible in bulk materials. In this section, we review the material systems, microscopic mechanisms, experimental realizations, and emergent phenomena of S/M heterostructures, emphasizing their role in generating spin-triplet superconductivity, nonreciprocal transport, and topological boundary excitations.

**3.1 Two-dimensional magnets as proximity building blocks**

The discovery of intrinsic magnetism in atomically thin vdW crystals has fundamentally reshaped the landscape of S/M heterostructures. Unlike conventional bulk magnets, which often suffer from lattice mismatch, surface roughness, and interfacial disorder, two-dimensional magnets provide atomically flat, chemically inert, and dangling-bond-free interfaces. These characteristics enable the fabrication of exceptionally clean S/M junctions, in which proximity effects can be studied with unprecedented precision and minimal extrinsic complications.

A growing family of layered magnetic materials[143-150]—including $Cr_2Ge_2Te_6$ (CGT), $CrI_3$, $CrBr_3$, and $Fe_3GeTe_2$ (FGT)—exhibits robust long-range magnetic order down to the few-layer or even monolayer limit. CGT is the first experimentally confirmed intrinsic vdW ferromagnet in the two-dimensional limit[143] (Fig. 5(a)–(b)), marking a milestone that effectively launched the field of two-dimensional magnetism. Subsequently, chromium trihalides $CrX_3$ (X = I, Br) attract considerable attention owing to their pronounced layer-number–dependent magnetic behavior[144] (Fig. 5(c)). For example, the magnetic ground state of $CrI_3$ evolves from ferromagnetism in the monolayer (Fig. 5(d)), to antiferromagnetic interlayer coupling in the bilayer, and back to ferromagnetism in the trilayer and bulk, giving rise to highly tunable magnetic ground states in the two-dimensional limit. In parallel, bulk and few-layer FGT exhibit itinerant ferromagnetism with a gate-tunable Curie temperature[145] (Fig. 5(e)–(f)) that significantly exceeds that of most other two-dimensional magnetic materials. These vdW magnets host a wide variety of magnetic ground states, ranging from Ising-type out-of-plane ferromagnetism to in-plane and canted spin configurations. Importantly, their magnetic anisotropy and ordering temperatures can be systematically tuned through thickness control,[144] electrostatic gating,[58,59] strain engineering,[68] and chemical intercalation.[146,147,149] This high degree of tunability provides exceptional control over the magnitude and orientation of the interfacial exchange field experienced by an adjacent superconductor.

Such controllability is particularly advantageous for proximity-coupled superconducting systems. For instance, in thin flakes of FGT, the application of a uniaxial tensile strain as small as ~0.3% can enhance the coercive field by more than 150%[68] (Fig. 5(g)–(h)), reflecting the extreme sensitivity of the magnetic anisotropy energy to lattice deformation (Fig. 5(i)). Such strain-controlled magnetism directly translates into a tunable exchange interaction at the S/M interface, allowing the amplitude, symmetry, and spatial profile of induced superconducting correlations to be actively manipulated. In this sense, two-dimensional magnets function not merely as passive sources of exchange fields, but as reconfigurable control elements for engineering superconducting quantum states.

Beyond static magnetic properties, the atomically thin geometry of vdW magnets strongly enhances interfacial SOC, particularly when combined with superconductors possessing intrinsically strong SOC or with heavy-element substrates. The coexistence of exchange fields and SOC at inversion-asymmetric interfaces provides a natural setting for singlet–triplet conversion, enabling the emergence of odd-frequency and equal-spin triplet Cooper pairs that can penetrate deep into the ferromagnetic layer.[151,152] Moreover, interfacial SOC induces spin–momentum locking and non-centrosymmetric superconductivity, which are essential ingredients for realizing nonreciprocal superconducting transport and magnetoelectric effects.[15,153-156] Therefore, two-dimensional magnetic materials establish an unprecedented proximity element for S/M heterostructures, in which magnetic order, exchange fields, and spin–orbit interactions can be continuously and independently tuned. This unique combination lays the microscopic foundation for many of the emergent phenomena discussed in subsequent sections, including long-range triplet superconductivity, superconducting diode effects, and topological boundary excitations.[15,28,29,152,154,157-160]

**3.2 Proximity effect in S/M heterostructures**

At the heart of S/M heterostructures lies the proximity-induced coupling between superconducting condensates and magnetic exchange fields. When a conventional spin-singlet superconductor is interfaced with a ferromagnet, Cooper pairs penetrating into the magnetic region experience an effective exchange field that lifts spin degeneracy. As a result, the superconducting pair amplitude acquires a finite center-of-mass momentum and oscillates spatially, giving rise to characteristic phenomena such as the suppression and revival of supercurrents[21] and the emergence of 0–π transitions in Josephson junctions.[161-163] Crucially, when strong interfacial SOC and broken inversion and time-reversal symmetries are present—as is generically the case in 2D vdW heterostructures—spin-singlet Cooper pairs can be converted into equal-spin triplet correlations. Unlike singlet pairs, these triplet pairs are immune to pair breaking by exchange fields and can propagate over long distances inside ferromagnets, giving rise to long-range, spin-polarized supercurrents.[164,165] This singlet–triplet conversion mechanism constitutes the microscopic foundation of superconducting spintronics, enabling dissipationless spin transport without accompanying Joule heating. Theoretical studies further predict that magnetic proximity and Rashba-type SOC at S/M interfaces can stabilize exotic pairing symmetries and magnetoelectric effects.[15,17,18,21,156,166-168] The experimental realization of proximity-induced superconductivity in atomically thin ferromagnets has provided compelling evidence of the unconventional nature of S/M coupling.

Direct experimental evidence for unconventional proximity coupling has been obtained in several two-dimensional S/M heterostructures. A representative example is the NbSe$_2$/FGT vdW heterostructure, where Hu et al.[169] demonstrated the coexistence of intrinsic ferromagnetism and proximity-induced superconductivity within the FGT layer. Resistance–temperature measurements reveal a clear superconducting transition in ultrathin FGT flakes (~4 nm), while thicker FGT layers exhibit a pronounced resistance upturn near the superconducting transition, reflecting the finite penetration depth of induced superconductivity (Fig. 6(a)). These results provide a transport evidence that superconductivity can coexist with ferromagnetism in atomically thin magnetic layers. STS further corroborates the coexistence of superconductivity and magnetism at the local level. In CrBr$_3$/NbSe$_2$ heterostructures, Kezilebieke et al.[170] observed a superconducting gap in the differential conductance spectrum measured at the center of CrBr$_3$ islands (Fig. 6(b)), demonstrating that superconducting correlations are induced directly inside the magnetic layer. Under an applied out-of-plane magnetic field, spatially resolved d$I$/d$V$ maps reveal a highly ordered hexagonal vortex lattice extending seamlessly across regions covered by CrBr$_3$ and bare NbSe$_2$ (Fig. 6(c)), indicating that the induced superconducting state remains phase coherent and compatible with the underlying magnetic order. Beyond inducing superconductivity in magnetic layers, proximity coupling can even enhance superconducting pairing in the presence of magnetism. In a MnTe/Bi$_2$Te$_3$/Fe(Te,Se) trilayer heterostructure, Ding et al.[29] reported a significantly-enhanced superconducting gap of approximately 2.9 meV on the surface of monolayer MnTe—substantially larger than the gap of the underlying Fe(Te,Se) superconductor (Fig. 6(d)). Notably, this enhancement disappears for bilayer MnTe samples, where only a conventional superconducting gap is observed. This striking layer-dependent behavior highlights the crucial role of interfacial magnetism and SOC in reshaping superconducting pairing at the atomic limit.

The cooperative interplay between proximity effects and strong SOC can further reconstruct magnetic anisotropy and even induce magnetic order inside superconductors. In NbSe$_2$/V$_5$Se$_8$ heterostructures, anomalous Hall effect (AHE) measurements reveal a dramatic evolution of magnetic behavior compared to V$_5$Se$_8$/TiSe$_2$. While V$_5$Se$_8$/TiSe$_2$ exhibits weak anisotropy consistent with a quasi-Heisenberg magnetic state, NbSe$_2$/V$_5$Se$_8$ displays a square-shaped hysteresis loop under out-of-plane magnetic fields, indicating that V$_5$Se$_8$ is switched from the Heisenberg ferromagnet with weak anisotropy to the Ising ferromagnet with strong anisotropy in NbSe$_2$/V$_5$Se$_8$[17] (Fig. 6(e)). Besides, angle-dependent AHE measurements further show anomalous deviations from the expected cos(θ) scaling[18] (Fig. 6(f)), suggesting the emergence of a proximity-induced ferromagnetic ground state within the superconducting NbSe$_2$ layer, driven by interfacial exchange coupling and SOC. Taken together, these experimental advances establish that proximity effects in two-dimensional S/M heterostructures generate genuinely new quantum states, rather than trivial replicas of bulk superconductivity or magnetism. The cooperative and competitive interplay among exchange fields, superconducting pairing, and spin–orbit interactions give rise to emergent interfacial phases characterized by long-range triplet correlations, enhanced or modified superconducting gaps, reconstructed magnetic anisotropy, and magnetized superconducting ground states. These phenomena provide the microscopic basis for the nonreciprocal transport, superconducting diode effects, and topological excitations discussed in the following sections.

**3.3 Symmetry breaking and nonreciprocal transport**

A defining consequence of interfacial symmetry breaking in S/M heterostructures is the emergence of asymmetric superconducting transport, in which the response of the system depends on the direction of charge flow. When inversion symmetry is broken at the interface and time-reversal symmetry is lifted by magnetic proximity effect, the superconducting free energy acquires momentum-asymmetric terms. As a result, Cooper-pair depairing becomes direction dependent, giving rise to nonreciprocal critical currents and rectification of supercurrents—phenomena collectively referred to as the SDE. From this perspective, nonreciprocal superconducting transport and extreme magnetoresistance can be viewed as closely related manifestations of the same underlying physics: an asymmetric coupling between superconductivity, magnetism, and spin–orbit interaction at vdW interfaces. In $NbSe_2$/CrSBr heterostructures,[153] for example, a magnetic-field-driven antiferromagnetic–ferromagnetic transition in CrSBr generates highly localized out-of-plane stray fields near the magnetic edges (Fig. 7(a)). These stray fields penetrate into the neighboring $NbSe_2$ and selectively suppress superconductivity over nanometer length scales, switching the system between a zero-resistance and a finite-resistance state. This abrupt superconducting on–off transition produces an effectively infinite magnetoresistance, illustrating how magnetic textures can act as local phase-control elements for superconducting transport. Notably, the combination of such a giant magnetoresistance with a small resistance–area product enables ultrahigh-density superconducting magnetic memory circuits (Fig. 7(b)). Early realizations of superconducting diodes in vdW S/M heterostructures relied on external magnetic fields to explicitly break time-reversal symmetry. In $NbSe_2$/$CrPS_4$ bilayers and $CrPS_4$/$NbSe_2$/$CrPS_4$ trilayer spin-valve structures,[154] magnetic proximity induces sizable critical-current asymmetry under modest applied fields. The enhanced nonreciprocity observed in the trilayer geometry reflects the increased magnetic asymmetry imposed by the two magnetic layers, demonstrating that vdW stacking offers a flexible route to engineer the strength and polarity of superconducting rectification. Conversely, the nonreciprocal transport can also be suppressed depending on the interfacial magnetic environment. For example, Che et al. reported that in $Fe_2O_3$/$NbSe_2$ heterostructures, the asymmetric superconducting response observed in pristine $NbSe_2$ is significantly reduced due to competing magnetic effects at the interface.[171] This highlights that interfacial magnetic textures can either enhance or diminish superconducting rectification, emphasizing the tunability of the SDE through careful interface engineering.

A major conceptual advance was achieved with the realization of field-free nonreciprocal superconductivity enabled by intrinsic interfacial symmetry breaking. In FGT/$NbSe_2$ heterostructures,[15] the strong SOC of Ising superconducting $NbSe_2$, combined with the interfacial exchange field from the vdW ferromagnet, breaks time-reversal symmetry without any external magnetic field. As a consequence, robust nonreciprocal supercurrents persist at zero field, with diode efficiencies comparable to or exceeding those of field-driven systems. This establishes that magnetic proximity alone, when coupled to strong SOC, is sufficient to generate intrinsic superconducting rectification. Beyond passive nonreciprocity, these systems further enable active and programmable control of superconducting asymmetry (Fig. 7(c)). In FGT/NbSe$_2$ devices, the polarity of the superconducting diode can be deterministically reversed by electrical current pulses, allowing all-electrical switching between opposite diode states. This reconfigurability enables the implementation of logic functionalities, such as XOR operations (Fig. 7(d)), and the faithful emulation of neuron-like response characteristics inspired by cortical information processing (Fig. 7(e)). Such behavior points to a magnetoelectric-like coupling between the superconducting condensate and the interfacial magnetic configuration, elevating the superconducting diode from a symmetry-derived effect to a functional, reconfigurable quantum device element. These developments establish S/M vdW heterostructures as a versatile platform for engineering asymmetric superconducting responses, ranging from infinite magnetoresistance to field-free, electrically programmable superconducting diodes. By introducing symmetry breaking directly at the interface, these systems open new pathways toward dissipationless superconducting electronics with functionalities unattainable in conventional superconductors.

### 3.4 Topological superconductivity and Majorana boundary modes

S/M heterostructures constitute one of the most promising platforms for engineering topological superconductivity in two dimensions. When a conventional spin-singlet superconductor is coupled to a magnetic layer in the presence of strong SOC, the combined system can effectively emulate a spinless *p*-wave superconductor. In such a regime, the superconducting state may acquire a nontrivial topological invariant, supporting MZMs localized at sample edges, magnetic domain walls, and vortex cores. Over the past decade, experimental signatures consistent with Majorana physics have been reported in a variety of hybrid architectures, including ferromagnetic insulator/superconductor interfaces, magnetic topological insulator/superconductor heterostructures, and engineered vdW magnetic–superconducting systems. Observations such as zero-bias conductance peaks, anomalous Josephson responses, and spatially localized zero-energy states detected by scanning tunneling microscopy and spectroscopy have provided compelling, though not yet definitive, evidence for topological boundary excitations. Establishing unambiguous Majorana signatures and disentangling them from trivial bound states remain central challenges in the field.

A typical realization of a designer two-dimensional topological superconductor was reported in CrBr$_3$/NbSe$_2$ heterostructures.[28] In this system, structural inversion symmetry breaking at the CrBr$_3$/NbSe$_2$ interface generates strong Rashba SOC, while intrinsic magnetism in CrBr$_3$ and superconductivity in NbSe$_2$ jointly provide all essential ingredients—magnetism, superconducting pairing, and SOC—for topological superconductivity. Differential conductance spectra acquired along the edges of CrBr$_3$ magnetic islands exhibit pronounced zero-bias features consistent with chiral Majorana modes. Spatially resolved d$I$/d$V$ line cuts across the island boundary (Fig. 8(a)) further reveal localized zero-energy states confined to the magnetic edge. The experimentally observed spectroscopic features are well reproduced by theoretical calculations of the local density of states, lending strong support to the interpretation in terms of a two-dimensional topological superconducting phase hosted by a monolayer ferromagnet/superconductor interface. Subsequent high-resolution STM/STS measurements by Feng et al. provided a critical reinvestigation of the $CrBr_3$/$NbSe_2$ heterostructure.[172] They showed that the emergence of edge states is closely correlated with lattice reconstruction at the edges of $CrBr_3$ islands. Three representative types of tunneling spectra were observed at the edges: a fully developed superconducting gap, a zero-energy conductance peak (ZECP), and in-gap bound states. Moreover, the zero-energy peak rapidly splits into symmetric side peaks with increasing tunneling transparency. These behaviors are inconsistent with robust topological Majorana edge modes and instead point to conventional Yu–Shiba–Rusinov states associated with localized magnetic moments, underscoring the critical role of interfacial coupling and edge reconstruction.

Complementary evidence for topological superconductivity and MZM has been reported in more complex vdW heterostructures that deliberately combine superconductivity, magnetism, and strong SOC. In a high-quality MnTe/$Bi_2Te_3$/Fe(Te,Se) trilayer system,[29] Ding et al. realized a platform in which conventional *s*-wave superconductivity is supplied by Fe(Te,Se), strong SOC arises from the topological insulator $Bi_2Te_3$, and thickness-dependent magnetism is provided by MnTe. STS measurements performed on monolayer (1-UC) MnTe reveal several striking features, including a robust zero-energy conductance peak and multiple discrete in-gap states(Fig. 8(b)). The coexistence of these spectroscopic signatures is consistent with the emergence of a nontrivial topological superconducting state and localized MZMs at the surface. Beyond isolated Majorana bound states, recent theoretical proposals suggest that engineered magnetic textures[173-175]—such as domain walls and skyrmions—in two-dimensional ferromagnets can host extended networks of topological superconducting channels. Such networks offer a potential route toward the controlled braiding and manipulation of non-Abelian quasiparticles in fully planar geometries. These advances position two-dimensional S/M heterostructures as a central building block for realizing topological superconductivity and for developing scalable platforms for topological quantum information processing.

### 3.5 Magnetic Josephson junctions and phase engineering

Magnetic Josephson junctions based on 2D S/M heterostructures provide a versatile and conceptually powerful platform for engineering superconducting phase, pairing symmetry, and current flow in a programmable manner. When a ferromagnetic layer is inserted between two superconducting electrodes, the exchange field modifies the superconducting current–phase relation, enabling phase shifts that go beyond the conventional 0-junction behavior. Depending on the magnetic configuration, interfacial SOC, and junction geometry, such systems can host π-junctions and φ₀-junctions with spontaneous phase offsets in the absence of external phase bias. A hallmark of magnetic Josephson junctions is the emergence of nontrivial phase ground states originating from the spatially inhomogeneous magnetic texture and proximity-induced superconducting correlations. In atomically thin magnetic-insulator Josephson junctions based on NbSe$_2$/Cr$_2$Ge$_2$Te$_6$/NbSe$_2$ heterostructures, Idzuchi et al.[162] demonstrated the realization of junctions whose ground-state phases deviate from both 0 and π. By comparing switching currents obtained under different bias sweep histories, the authors extracted two degenerate phase offsets, neither corresponding to an integer multiple of π (Fig. 8(c)). This behavior is understood as arising from the coexistence and interference of 0- and π-junction segments induced by heterogeneous magnetic domains and tilted spins in the Cr$_2$Ge$_2$Te$_6$ barrier (Fig. 8(d)). Such domain-mediated interference effectively realizes a φ$_0$-junction with a doubly degenerate phase ground state, highlighting how magnetic microstructure can be directly encoded into superconducting phase.

Beyond static phase shifts, vdW Josephson junctions enable dynamic and reconfigurable control of superconducting phase and supercurrent. The atomically sharp interfaces and high material cleanliness inherent to vdW assembly strongly enhance proximity effects, allowing unconventional superconducting correlations to propagate over long distances. A demonstration was reported by Hu et al.,[165] who fabricated lateral S/F/S Josephson junctions using the vdW ferromagnet FGT as the magnetic barrier and NbSe$_2$ as superconducting electrodes. Remarkably, supercurrents were observed to propagate across FGT over distances exceeding 300 nm—far beyond the decay length expected for spin-singlet pairs. This long-range Josephson coupling provides compelling evidence for the generation of equal-spin triplet supercurrents, which are robust against exchange fields and constitute a central building block of superconducting spintronics. Quantum interference measurements in these junctions further revealed Fraunhofer patterns shifted away from zero magnetic field, reflecting the intrinsic magnetization of the ferromagnetic barrier and its direct coupling to the Josephson phase. Even more intriguingly, the supercurrent density is found to be strongly localized near the surfaces or edges of the ferromagnet, forming so-called skin Josephson currents (Fig. 8(f)). This highly nonuniform current distribution represents a novel transport regime enabled by reduced dimensionality, strong magnetism, and interfacial proximity effects, and has no direct analogue in conventional bulk Josephson junctions.

Magnetic Josephson junctions in 2D S/M heterostructures transform magnetism from an antagonistic perturbation into an active control knob for superconducting phase and pairing symmetry. By enabling long-range triplet supercurrents, nontrivial phase offsets, and spatially engineered supercurrent flow, these systems establish a foundation for superconducting phase batteries, reconfigurable Josephson circuits, and hybrid quantum devices. More broadly, the concepts developed here naturally extend to Josephson junctions incorporating topological and strongly correlated materials, providing a key bridge between proximity-engineered superconductivity and the topological superconducting phenomena discussed in the following sections.

## 4. Superconductor/topological material heterostructures

Superconductor/topological-material (S/T) heterostructures constitute a rapidly expanding arena in which superconductivity intertwines with topological electronic structures, SOC and crystalline symmetry. The central concept is that superconducting correlations, when injected into topological surface or interface states, do not remain conventional but reorganize into new pairing configurations governed by the underlying topological or symmetry-protected degrees of freedom. The resulting hybrid superconducting states often display mixed-parity pairing, finite-momentum condensates, nonreciprocal responses, phase-biased Josephson behavior, and highly tunable quasiparticle excitations. These emergent effects provide not only a platform for exploring unconventional superconductivity but also fertile ground for realizing topological quasiparticles such as Majorana bound states[42,176-178] and for engineering functional superconducting devices beyond the paradigm of conventional Josephson physics. In this section, we review the microscopic mechanisms, representative material systems, experimental signatures, and emerging device concepts associated with S/T heterostructures. Special emphasis is placed on how symmetry, topology, and quantum geometric properties shape the interfacial superconducting state and how these features give rise to a spectrum of measurable phenomena and functional capabilities.

**4.1. Microscopic mechanisms at S/T interfaces**

At the heart of S/T heterostructures lies the conversion of conventional Cooper pairs into interfacial pairing amplitudes that reflect the symmetry and spin texture of the topological material. Because topological surface states exhibit spin–momentum locking or Rashba-type spin–orbit fields, the induced pairing naturally acquires mixed singlet–triplet character.[179-184] In the Fu–Kane framework,[45] for example, $s$-wave proximity coupling to a Dirac surface state generates an effective p-wave component capable of hosting Majorana modes. More generally, interfacial superconductivity arises from a combination of spin–orbit interactions, band inversion, and orbital hybridization, each contributing distinct channels for pairing conversion.[185-189] Symmetry considerations further dictate the nature of the emergent order. Breaking inversion symmetry at the interface favors parity mixing and allows finite-momentum Cooper pairing, which can manifest as helical superconductivity or lead to Josephson phase shifts in $\varphi_0$-junctions.[20,23,190,191] These effects could give rise to unconventional phenomena such as nonreciprocal Josephson responses and phase-bias generation.[23,192,193] Overall, these mechanisms establish a unified physical landscape for understanding how superconducting correlations evolve at topological interfaces. The induced pairing is therefore not simply inherited from the parent superconductor; rather, it emerges as a hybrid condensate shaped jointly by symmetry and topology. This framework provides the foundation for analyzing specific material platforms and interpreting a wide variety of proximity-induced superconducting phenomena.

**4.2. Topological insulator–superconductor interfaces**

Topological insulators (TIs), such as $Bi_2Se_3$, $Bi_2Te_3$, and $Sb_2Te_3$,[194] host gapless Dirac surface states with helical spin textures protected by time-reversal symmetry.

When such surface states are interfaced with a conventional s-wave superconductor, superconducting correlations penetrate into the topological layer via the superconducting proximity effect. Owing to the spin-helical nature of the Dirac fermions, the induced pairing is fundamentally reshaped: within the Fu–Kane framework,[45] conventional s-wave pairing is effectively converted into a spinless p-wave channel. This conversion enables two-dimensional topological superconductivity that can host MZMs at vortices, magnetic domain boundaries, or sample edges.[195-198]

Ultrathin TI films provide an additional and highly effective degree of tunability for proximity-induced superconductivity. In these systems, robust interfacial superconductivity can be stabilized in few-quintuple-layer (QL) systems when coupled to a superconductor. Molecular-beam-epitaxy-grown $Bi_2Se_3$/$NbSe_2$ and $Bi_2Te_3$/$NbSe_2$ heterostructures have therefore emerged as prototypical platforms, combining atomically sharp interfaces, high crystalline quality, and precise thickness control. In these systems, scanning tunneling microscopy and spectroscopy (STM/STS) directly visualize proximity-induced superconducting gaps as well as vortex-core bound states, providing spatially resolved access to the interfacial superconducting order. A seminal experimental demonstration of the coexistence between superconductivity and topological surface states is reported by Wang et al. in $Bi_2Se_3$/$NbSe_2$ heterostructures. Low-temperature STM/STS measurements[185] (Fig. 9(a)) reveal a well-defined superconducting gap on the surface of $Bi_2Se_3$, whose magnitude gradually decreases with increasing film thickness (Fig. 9(b)), consistent with expectations for a proximity-induced order parameter. Complementary ARPES measurements (Fig. 9(c)) further show that topological Dirac surface states emerge when the $Bi_2Se_3$ thickness reaches approximately six quintuple layers, with a clearly resolved Dirac point located about 0.45 eV below the Fermi level. The simultaneous observation of a superconducting gap and intact Dirac surface states established that superconductivity and nontrivial band topology can coexist within a single, well-controlled heterostructure, thereby laying a crucial experimental foundation for realizing Majorana physics in the S/T heterostructures.

Beyond conventional superconducting proximity, S/T heterostructures also enable the topological proximity effect,[199] whereby the nontrivial band topology of a TI is partially transferred into an adjacent superconductor. In Pb/$TlBiSe_2$ heterostructures, for example, ARPES measurements revealed Dirac-like electronic states at the Pb–vacuum interface that originate from the buried TI substrate rather than from Pb itself (Fig. 9(d)). Spin-selective hybridization between TI Dirac states and Pb quantum-well states stabilizes a topologically nontrivial electronic structure within the superconducting film, offering a complementary route toward engineering topological superconductivity that does not rely solely on inducing superconductivity in the TI layer. The MZMs at the vortex core is a hallmark feature of topological superconductivity and has been investigated in several S/T systems using STM/STS. However, identifying MZMs remains experimentally challenging, as conventional zero-bias conductance peaks can be mimicked by low-energy Caroli–de Gennes–Matricon states in vortex cores. An effective strategy to enhance selectivity exploits the spin structure of Majorana modes: the Majorana wave function at the vortex center is predicted to be fully spin polarized, rendering zero-energy Andreev reflection intrinsically spin selective. Spin-polarized STM/STS therefore provides a more stringent probe of Majorana physics. This approach was experimentally demonstrated by Sun et al.[198] in $Bi_2Te_3$/$NbSe_2$ heterostructures (Fig. 9(e)), where spin-polarized STM/STS revealed a pronounced spin-dependent enhancement of the zero-bias conductance localized at vortex centers. These observations significantly strengthen the identification of the zero-energy bound states as MZMs and establish S/T heterostructures as experimentally accessible platforms for probing topological superconductivity.

### 4.3. Topological semimetal–superconductor interfaces

Topological semimetals offer a qualitatively distinct route to realizing interfacial topological superconductivity, fundamentally different from that based on topological insulators. In Dirac and Weyl semimetals,[183,200-202] the bulk band structure hosts symmetry-protected gapless nodes, accompanied by open Fermi-arc surface states rather than closed Dirac cones. These extended surface manifolds, together with strong SOC and reduced symmetry, enable proximity-induced superconducting correlations that are inherently anisotropic, and sensitive to band topology.

Type-II Weyl semimetals, exemplified by $WTe_2$- and $MoTe_2$-based compounds,[181,203-207] provide a rich playground. Their strongly tilted Weyl cones and highly anisotropic Fermi surfaces break Lorentz invariance and fundamentally reshape the superconducting proximity effect. In $WTe_2$ (Fig. 10(a)), strong SOC, band inversion, and lattice distortion cooperate to generate a nontrivial bulk topology[181] accompanied by helical boundary modes (Fig. 10(b)). In addition to its topological band structure, $WTe_2$ exhibits a range of hallmark transport phenomena, including pronounced charge–spin conversion[205] driven by Berry curvature and spin texture (Fig. 10(c)), extremely large and non-saturating positive magnetoresistance[204] governed by carrier compensation (Fig. 10(d)), and the chiral anomaly manifested as negative longitudinal magnetoresistance[203] (Fig. 10(e)). These properties highlight $WTe_2$ as a highly tunable quantum material in which topology, geometry, and transport are intrinsically intertwined. When proximitized by an *s*-wave superconductor, $WTe_2$ gives rise to unconventional superconducting states that directly reflect its topological electronic structure. In $WTe_2$/$NbSe_2$ vdW heterostructures, proximity-induced superconductivity penetrates deep into the semimetal, indicating efficient Cooper-pair injection despite the absence of a bulk gap. Transport and spectroscopic measurements reveal long coherence lengths and unconventional subgap structures, signaling a departure from conventional proximity superconductivity. A particularly striking realization is the superconducting quantum spin Hall effect in monolayer $WTe_2$/$NbSe_2$ heterostructures. Low-temperature scanning tunneling microscopy and spectroscopy directly resolve enhanced local density of states (Fig. 11(a)) at the $WTe_2$ edges,[188] confirming the presence of quantum spin Hall edge channels. More importantly, a proximity-induced superconducting gap is observed within these edge states (Fig. 11(b)-(c)), providing compelling evidence for the coexistence of superconductivity and topologically protected helical transport. This achievement establishes a solid-state platform in which one-dimensional topological edge modes are coherently proximitized, opening pathways toward Majorana-based boundary excitations.

Beyond edge superconductivity, proximity effects in WTe$_2$ can generate anomalous subgap spectral features rooted in the intrinsic electronic structure of the Weyl semimetal. In WTe$_2$/NbSe$_2$ heterostructures, differential resistance measurements (Fig. 11(d)) reveal multiple subgap peaks in addition to the conventional coherence features.[208] Theoretical modeling of the density of states (Fig. 11(e)-(f)) reproduces these anomalies and indicates that superconducting WTe$_2$ develops secondary coherence peaks at energies significantly smaller than the parent NbSe$_2$ gap. This behavior reflects a multi-step Cooper-pair depairing process associated with momentum-selective pairing on Weyl-derived states, highlighting the nontrivial nature of proximity-induced superconductivity in topological semimetals. Topological semimetal–superconductor interfaces also exhibit pronounced nonreciprocal and diode-like superconducting responses. A representative example is the giant Josephson diode effect (Fig. 11(g)) observed in NiTe$_2$/Nb junctions,[23] where the critical current becomes strongly asymmetric with respect to the current direction under an in-plane magnetic field. This nonreciprocity originates from finite-momentum Cooper pairing (Fig. 11(h)-(i)) enabled by spin–momentum-locked surface states and broken inversion symmetry. Such effects underscore the potential of topological semimetal–superconductor hybrids for realizing superconducting devices with intrinsic rectification and directionality. Overall, topological semimetal–superconductor interfaces highlight a unique interplay between superconductivity, spin–momentum locking, and band topology. Compared with topological insulator platforms, semimetal-based heterostructures offer enhanced tunability, anisotropic pairing, and access to gapless bulk and boundary states, enabling a broad spectrum of exotic superconducting phenomena and functional quantum device concepts.

## 5. Superconductor/superconductor (S/S) heterostructures

Compared with S/M or S/T heterostructures, superconductor/superconductor (S/S) interfaces offer a uniquely clean platform in which the Josephson effect itself can actively generate emergent quantum states, rather than merely probing phase coherence. In particular, twistronics—the deliberate introduction of a relative twist angle between two identical or similar superconducting layers—provides a continuously tunable degree of freedom that controls interlayer coupling, momentum-space matching, and moiré superlattice potentials. When combined with the atomic precision of vdW assembly, which enables high-quality interfaces and precise control over crystallographic orientation, S/S heterostructures allow interlayer hybridization, phase coherence, and symmetry constraints to be deliberately engineered. These capabilities transform superconducting interfaces into active quantum materials, in which the superconducting order can be tailored to realize unconventional Josephson responses, higher-order tunneling processes, novel pairing symmetries and nonreciprocal transport.

### 5.1. Twist-engineered Josephson physics in S/S heterostructures

Twist engineering has recently emerged as a powerful strategy for creating novel superconducting and topological phenomena in S/S heterostructures. By rotating two superconducting layers relative to one another, long-wavelength moiré superlattices are formed, introducing momentum mismatch and spatial modulation of the interlayer coupling. Even when the individual superconductors are well understood in isolation, a finite twist angle can fundamentally reshape the Josephson coupling, alter the effective pairing channels, and give rise to emergent collective states.

From a microscopic perspective, the Josephson current in twisted S/S junctions is no longer determined solely by the macroscopic phase difference, but also by the momentum-space overlap of the two condensates. Twist-induced momentum mismatch suppresses first-order tunneling processes and enhances the role of higher-order Josephson harmonics,[66,209,210] enabling finite-momentum Cooper pairing, phase-biased Josephson states, and unconventional current–phase relations. In systems lacking inversion or time-reversal symmetry, these effects naturally give rise to nonreciprocal superconducting transport, including superconducting diode behavior and directional Josephson responses.[63,65,211-213]

Compared with conventional composition-based heterostructure engineering, twist provides a uniquely clean and continuously tunable parameter space. It allows access to symmetry-breaking superconducting states without introducing chemical disorder or additional scattering channels. More broadly, twisted superconducting interfaces offer a versatile route toward designer Josephson circuits, tunable SQUID architectures, and programmable superconducting networks whose properties are encoded in geometry rather than material composition. Looking ahead, theoretical studies suggest that twist engineering can also induce emergent interfacial magnetism in a wide range of two-dimensional bilayer systems, including $NbSe_2$, $MoS_2$, and h-BN.[214] This magnetism arises from twist-driven spin splitting and enhanced exchange interactions, and can coexist or compete with superconductivity to produce unconventional topological and correlated phases, yet it remains largely unexplored experimentally. These insights point to an exciting opportunity: by combining moiré engineering with sensitive magnetometry and transport probes, one can realize tunable interfacial magnetic states in S/S heterostructures, establishing a direct pathway toward reconfigurable superconducting quantum devices.

## 5.2. Twisted cuprate Josephson junctions

Among twist-engineered S/S heterostructures, twisted cuprate superconductors—particularly BSCCO—occupy a unique position. As a prototypical high-temperature superconductor with a predominantly d-wave order parameter, BSCCO provides a stringent building block for testing how twist angle, pairing symmetry, and interlayer tunneling interplay in the two-dimensional limit. The extreme anisotropy and weak interlayer coupling of BSCCO further enable mechanical exfoliation and atomically clean vdW stacking, making it ideally suited for twist-controlled Josephson junctions.

Early theoretical proposals established twisted cuprate junctions as a phase-sensitive probe of superconducting order. For an ideal d-wave superconductor, the Josephson coupling is expected to vanish at a relative twist angle of 45°, reflecting the sign change of the order parameter under π/2 rotation, whereas a finite supercurrent would persist in the presence of an s-wave component or incoherent tunneling contributions[211] (Fig. 12(a)). More recent theoretical studies have proposed a qualitatively different scenario in which higher-order tunneling processes dominate near 45°, potentially stabilizing a chiral $d+id$ superconducting state. Such a state would spontaneously break time-reversal symmetry and manifest through anomalous temperature dependence of the critical current, modified Fraunhofer interference patterns, and fractional Shapiro steps, raising the intriguing possibility of topological superconductivity at elevated temperatures.

Experimentally, realizing twisted BSCCO Josephson junctions poses substantial challenges due to the intrinsic complexity of cuprate superconductors. Disorder, oxygen nonstoichiometry, and strong vortex pinning can obscure subtle interfacial effects, while ultrathin BSCCO flakes are highly sensitive to fabrication and environmental conditions. Recent advances—including inert-atmosphere assembly, graphene encapsulation, and low-temperature stacking—have nevertheless enabled the preservation of superconductivity down to the few–unit-cell and even half–unit-cell limit, opening a practical route toward systematic twist-angle control.

Despite these technical advances, experimental reports on twist-angle–dependent Josephson coupling remain conflicting. Several studies observe a strong suppression of the critical current near 45°[66] (Fig. 12(b)), accompanied by anomalous Fraunhofer interference patterns and fractional Shapiro steps, which have been interpreted as signatures of higher-order tunneling and possible chiral $d+id$ pairing. In contrast, high-precision measurements on highly uniform junctions report conventional Fraunhofer interference and well-defined Fiske resonances[65] (Fig. 12(c)), without the anomalies expected for dominant second-order tunneling processes. These results cast doubt on the robustness of a twist-stabilized $d+id$ superconducting state. Complementarily, other experiments observe sizable Josephson currents even near 45°, with $I_cR_n$ values comparable to those at small twist angles and a largely conventional temperature dependence[63] (Fig. 12(d)). Such behavior suggests the presence of an effective s-wave–like tunneling component or an incoherent admixture of pairing symmetries. At present, twisted BSCCO Josephson junctions therefore remain a paradigmatic yet controversial platform. Rather than delivering a definitive verdict on cuprate pairing symmetry, they highlight the extreme sensitivity of Josephson physics in cuprates to disorder, interface transparency, and fabrication protocols. Resolving whether twist engineering can genuinely stabilize a new topological superconducting phase—or instead reveals an unexpected resilience of s-wave–like Josephson coupling—remains an open and central challenge.

Beyond their fundamental implications, twisted cuprate junctions have also enabled direct control of nonreciprocal superconducting transport. In particular, several studies have reported a pronounced Josephson diode effect at specific twist angles. For example, Zhu et al.[215] observe sizable Josephson tunneling near 45° accompanied by a robust JDE persisting up to ~50 K in the overdoped regime (Fig. 12(e)), while Ghosh et al.[212] demonstrate a JDE surviving up to 77 K, which is attributed to the interaction between Josephson and Abrikosov vortices (Fig. 12(f)). More recently, Zhang et al.[213] report a quantized SDE operating entirely within the Cooper-paired manifold, exploiting quantized Shapiro steps to achieve digitized output with near-perfect diode efficiency above liquid-nitrogen temperature (Fig. 12(g)). Together, these results establish twist engineering as a powerful and conceptually new control knob for superconducting order, not only enabling stringent tests of unconventional pairing theories but also opening pathways toward high-temperature, nonreciprocal superconducting devices based on strongly correlated materials.

## 6. Conclusion and perspectives

In this review, we have provided an overview of emergent superconducting phenomena in 2D vdW heterostructures formed at interfaces between superconductors, magnetic materials, and topological systems. Enabled by atomically sharp interfaces, weak interlayer bonding, and extensive tunability, these hybrid structures provide an unprecedented platform for engineering superconductivity beyond the constraints of bulk materials. Through proximity effects, strong SOC, symmetry breaking, and twist-angle control, vdW heterostructures host a broad spectrum of unconventional superconducting states that are otherwise difficult or impossible to realize. A unifying principle across S/M, S/T, and S/S interfaces is the interfacial hybridization of electronic orders. Superconducting, magnetic, and topological orders can partially penetrate across atomically clean interfaces, generating emergent condensates with mixed symmetry, finite-momentum pairing, or spin-polarized character. Such interlayer coupling enables controlled coexistence and mutual reconstruction of superconductivity with magnetism, topology, and strong spin–orbit interaction, greatly expanding the accessible quantum phase space beyond that of the constituent layers.

In S/M heterostructures, proximity effects transform the conventional antagonism between superconductivity and magnetism into a versatile resource: magnetic order can be reconstructed or enhanced, while superconducting correlations penetrate magnetic layers as long-range spin-triplet pairs. These coupled orders give rise to magnetic Josephson junctions, programmable quantum phases, infinite magnetoresistance, and electrically controllable nonreciprocal transport, laying the foundation for superconducting spintronics and low-dissipation spin control. S/T heterostructures, by contrast, leverage superconducting and topological proximity effects to realize interfacial topological superconductivity. Such systems support MZMs, edge supercurrents, and spin-selective Andreev reflection, establishing vdW S/T platforms as prime candidates for topological quantum computation. Twist-angle engineering and moiré superlattices further enrich S/S heterostructures, providing a highly tunable control knob over Josephson coupling, phase coherence, and condensate symmetry. Even identical superconducting layers can host correlated electronic states and angle-dependent Josephson effects, demonstrating that twist can transform passive junctions into designer quantum materials.

Looking forward, vdW superconductor heterostructures provide a uniquely versatile platform for both exploring emergent quantum phenomena and developing next-generation quantum and neuromorphic devices. The combination of atomically sharp interfaces, weak interlayer bonding, and multiple control knobs—including electric fields, strain, twist angle, and thermal gradients—enables dynamic, reversible manipulation of superconducting, magnetic, and topological states. This tunability allows the realization of programmable nonreciprocal devices, superconducting diodes, and tunable Josephson networks, while also supporting neuromorphic superconducting architectures in which multiple stable quantum states and interfacial coupling can implement synaptic functionalities, in-memory computation, and ultralow-power signal processing. Twist engineering and moiré superlattices further enrich this landscape by reshaping interlayer hybridization, controlling phase coherence, and inducing novel correlated or topological superconducting states, offering a design-driven approach to both fundamental physics and functional device platforms. Beyond individual device concepts, vdW superconductor heterostructures establish a bridge between neuromorphic and quantum computing paradigms. Superconducting neuromorphic processors can function as high-speed, low-power classical co-processors for quantum systems, handling control, error correction, and real-time pre-processing, while quantum processors perform intrinsically quantum operations such as entanglement generation, variational optimization, and probabilistic sampling. By integrating tunable superconductivity, interfacial magnetism, and topological effects within a single materials platform, these heterostructures provide a unified foundation for exploring emergent quantum phases and for engineering hybrid computing architectures that combine classical, neuromorphic, and quantum functionalities. Collectively, this positions vdW superconductor heterostructures as a transformative arena for both fundamental discoveries and the realization of intelligent, energy-efficient, and reconfigurable quantum technologies.

## Acknowledgements

This work was supported in part by the National Key R&D Program of China under grant 2025YFA1411003, 2023YFF0718400 and 2023YFF1203600, the National Natural Science Foundation of China 12322407, 62034004, the Leading-edge Technology Program of Jiangsu Natural Science Foundation BK20232004, BK20233001, the Fundamental Research Funds for the Central Universities 14380227, 14380240, 14380247, 14380250, and 14380242, the Innovation Program for Quantum Science and Technology 2024ZD0300101. F.M. and S.-J.L. would like to acknowledge support from the AIQ Foundation and the e-Science Center of Collaborative Innovation Center of Advanced Microstructures. The microfabrication center of the National Laboratory of Solid State Microstructures (NLSSM) is acknowledged for their technique support.

# Figures

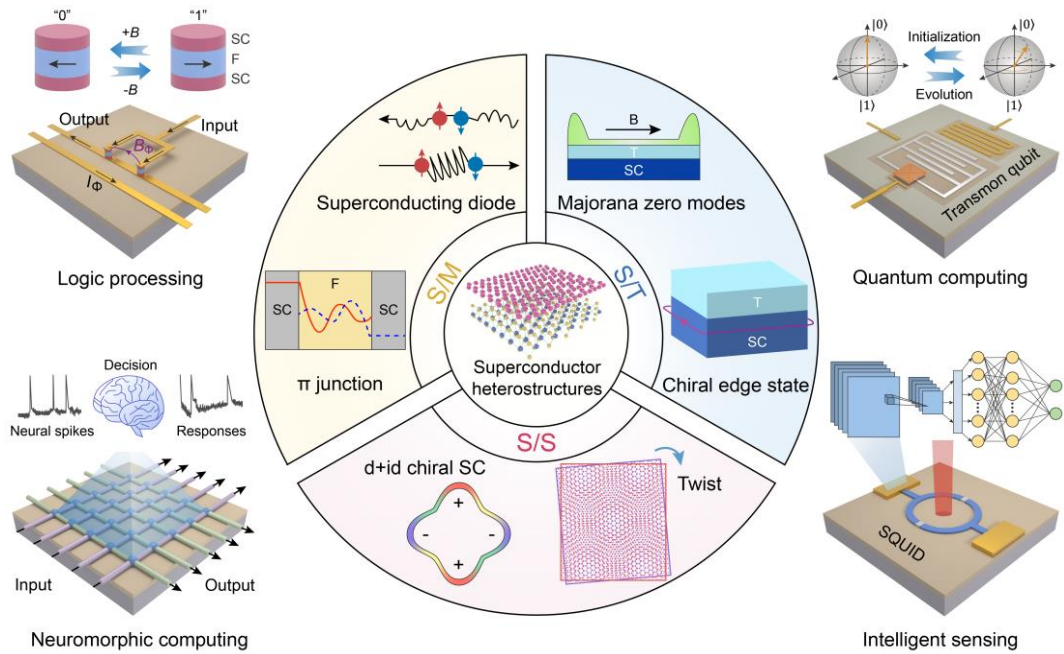

**Fig. 1.** Schematic illustration of three prototypical 2D superconductor heterostructures discussed in this review: superconductor/magnet (S/M), superconductor/topological material (S/T), and superconductor/superconductor (S/S) junctions. Interfacial proximity effects in these systems give rise to unconventional superconducting states, including spin-triplet pairing, topological superconductivity, nonreciprocal and chiral superconducting transport. These heterostructures provide a versatile platform for exploring emergent quantum phases and developing superconducting quantum devices and quantum intelligent computing and sensing.

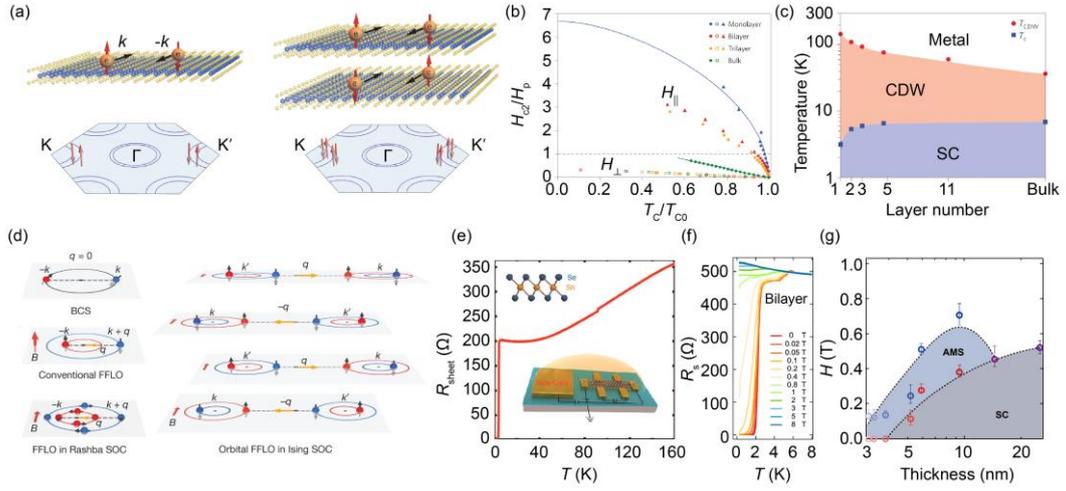

**Fig. 2.** Ising superconductivity and gate-induced superconductivity in representative 2D materials. (a) Illustration of spin–momentum locking in monolayer NbSe$_2$ and spin–layer locking in bilayer/bulk NbSe$_2$. (b) Normalized upper critical field $H_{c2}/H_p$ as a function of transition temperature $T_C/T_{C0}$ for NbSe$_2$ samples of differing thickness under both out-of-plane $H_\perp$ (open symbols) and in-plane $H_\parallel$ (filled symbols) magnetic fields. The dashed line corresponds to the Pauli paramagnetic limit. (c) Thickness–temperature phase diagram of NbSe$_2$. (d) Comparison of zero-momentum BCS pairing, finite-momentum FFLO pairing, and orbital FFLO states in Rashba and multilayer Ising superconductors. (e) Temperature-dependent resistance of a SnSe$_2$-EDLT device, showing a gate-induced superconducting transition. (f) Temperature-dependent sheet resistance of bilayer MoS$_2$ devices under perpendicular magnetic fields, demonstrating gate-induced superconductivity. (g) Phase diagram of thickness versus magnetic field in 2H-Ta$_2$S$_3$Se. Panels (a)-(b) are reproduced with permission from Ref.[77]. Panel (c) is reproduced with permission from Ref.[78]. Panel (d) is reproduced with permission from Ref.[80]. Panel (e) is reproduced with permission from Ref.[60]. Panel (f) is reproduced with permission from Ref.[57]. Panel (g) is reproduced with permission from Ref.[90].

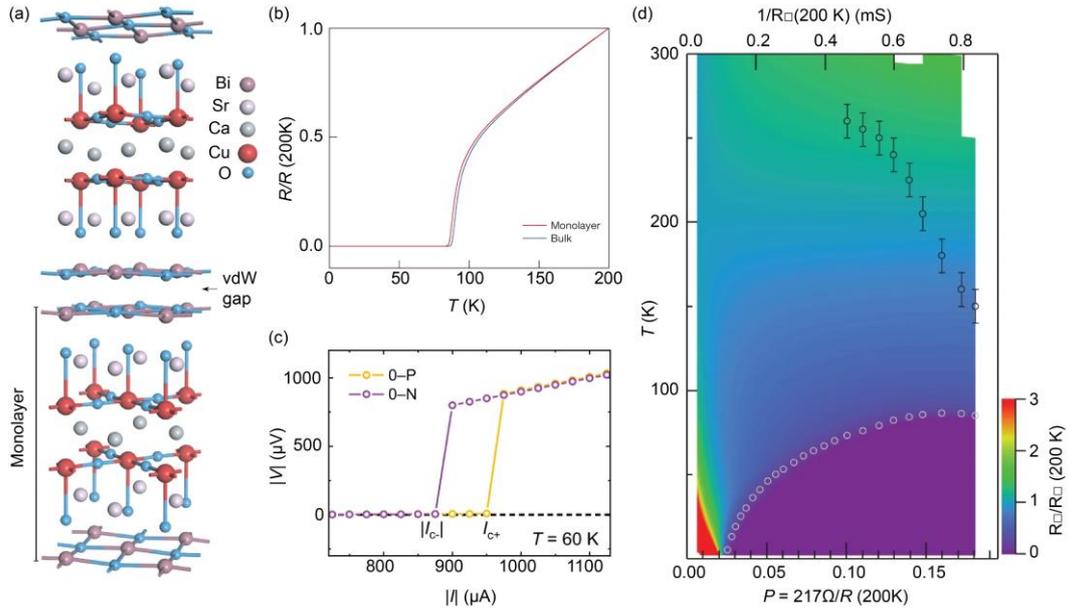

**Fig. 3.** Crystal structure and transport phenomena in Bi-2212. (a) Crystal structure of Bi-2212. A "monolayer" corresponds to a half unit cell along the out-of-plane direction, containing two $CuO_2$ planes. (b) Temperature-dependent resistance of a monolayer Bi-2212 device (red) compared with an optimally doped bulk crystal (blue). (c) Current–voltage characteristics at 60 K exhibiting nonreciprocal critical currents, with $I_c^+ \neq |I_c^-|$ for opposite current directions. (d) Resistance as a function of temperature and doping, illustrating the evolution of transport behavior across the phase diagram. Panels (a)(b)(d) are reproduced with permission from Ref.[110]. Panel (c) is reproduced with permission from Ref.[116].

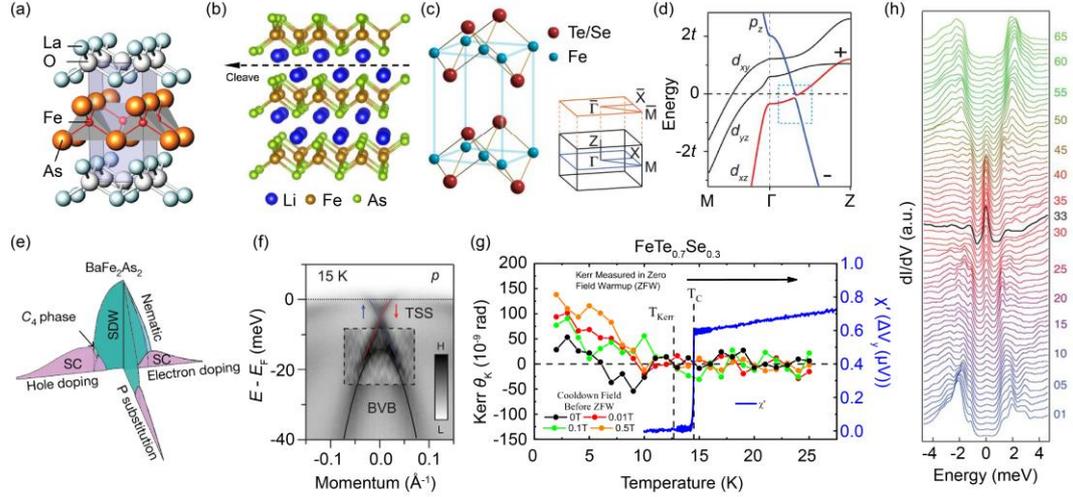

**Fig. 4.** Topology, symmetry breaking, and vortex states in iron-based superconductors. (a-c) Crystal structure of representative iron-based superconductors: LaOFeAs (a), LiFeAs (b), and Fe(Se,Te) (c). (d) First-principles band structure of Fe(Se,Te), highlighting a spin–orbit–coupling-induced gap associated with band inversion. (e) Schematic phase diagram of $BaFe_2As_2$, illustrating nematic order, spin-density-wave and $C_4$ magnetic phases, and superconductivity. (f) Band structure of Fe(Se,Te), showing a bulk hole-like valence band and a Dirac-cone–like topological surface state. (g) Zero-field warm-up Kerr rotation $\theta_K$ revealing spontaneous time-reversal symmetry breaking below $T_{Kerr}$, while bulk susceptibility $\chi^0$ marks the superconducting transition at $T_c$ without evidence of ferromagnetism. (h) Spatially resolved tunneling spectra across a vortex core, with the central spectrum highlighted. Panel (a) is reproduced with permission from Ref.[122]. Panel (b) is reproduced with permission from Ref.[127]. Panels (c)(d)(f) are reproduced with permission from Ref.[130]. Panel (e) is reproduced with permission from Ref.[133]. Panel (g) is reproduced with permission from Ref.[140]. Panel (h) is reproduced with permission from Ref.[141].

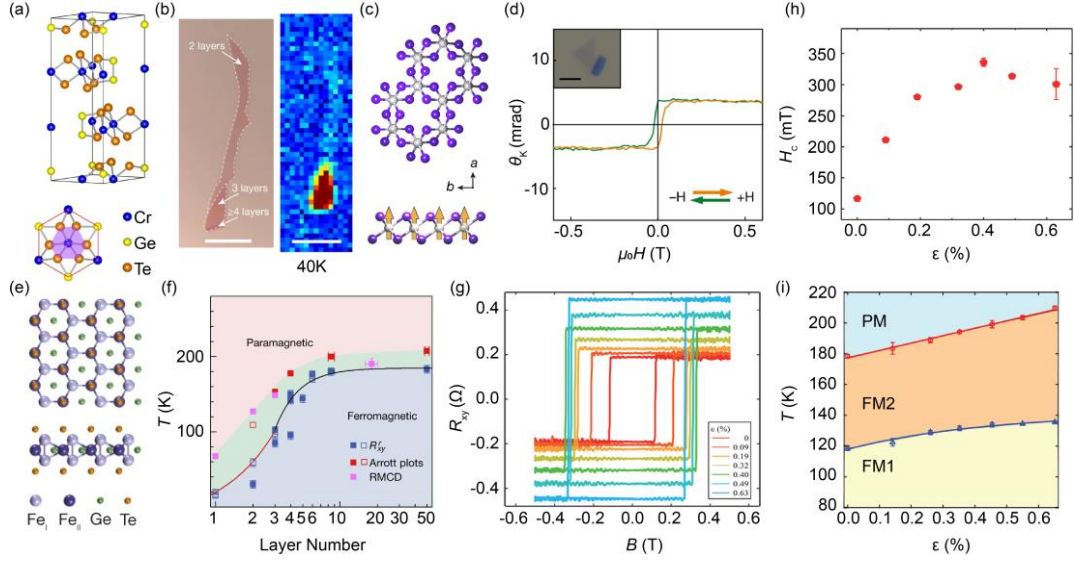

**Fig. 5.** Two-dimensional vdW magnets and tunable magnetic phases. (a) Crystal structure of $Cr_2Ge_2Te_6$ shown in side and top views. (b) Optical image of exfoliated $Cr_2Ge_2Te_6$ atomic layers on $SiO_2/Si$ (left) and corresponding Kerr rotation signal from a bilayer flake under an out-of-plane magnetic field at 40 K (right). (c) In-plane (top) and out-of-plane (bottom) atomic lattice of a monolayer $CrI_3$. (d) Polar MOKE signal of a $CrI_3$ monolayer. The inset shows an optical image of an isolated monolayer (the scale bar is 2 μm). (e) Atomic structure of monolayer FGT viewed along the [001] (top) and [010] (bottom) directions. (f) Layer-number–temperature phase diagram of FGT. (g,h) Strain-dependent anomalous Hall response and coercive field of FGT at 1.5 K. (i) Temperature–strain phase diagram of FGT, distinguishing paramagnetic (PM) and ferromagnetic phases with single-domain (FM1) and labyrinthine-domain (FM2) states. Panels (a)-(b) are reproduced with permission from Ref.[143]. Panels (c)-(d) are reproduced with permission from Ref.[144]. Panel (e)-(f) is reproduced with permission from Ref.[145]. Panels (g)-(i) are reproduced with permission from Ref.[68].

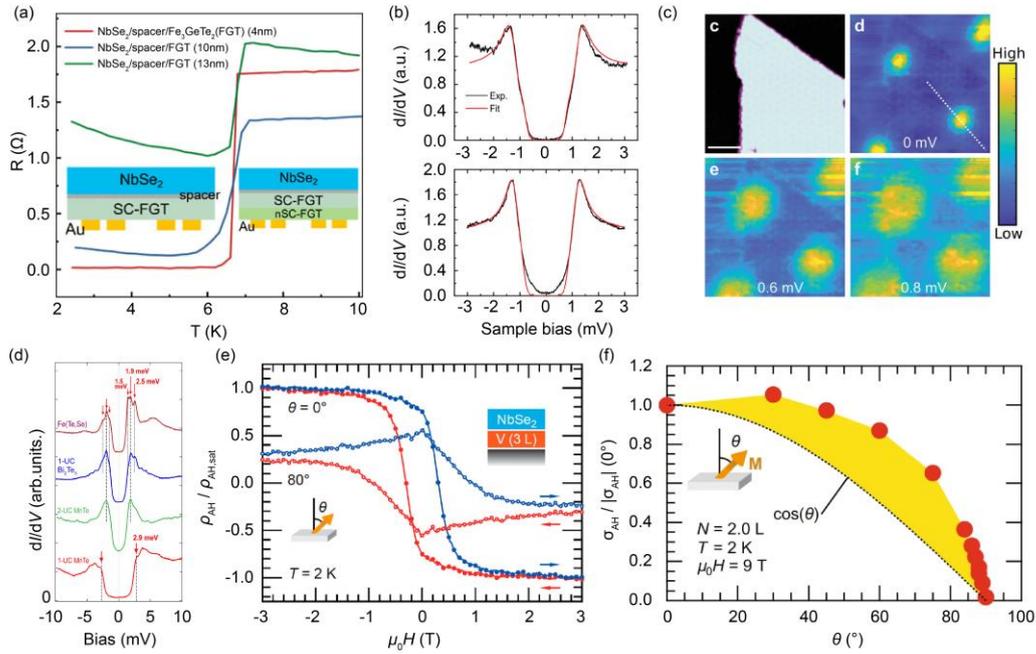

**Fig. 6.** Proximity effects and emergent phenomena in superconductor/magnet heterostructures. (a) Temperature-dependent resistance of NbSe$_2$/spacer/FGT with varying FGT thickness and fixed NbSe$_2$ thickness of ~20 nm. (b) Differential conductance spectra measured on bare NbSe$_2$ and on a magnetic CrBr$_3$ island, fitted by a double-gap s-wave BCS model, indicating proximity-modified superconductivity. (c) Vortex imaging on CrBr$_3$/NbSe$_2$ heterostructure, revealing magnetic modulation of superconducting vortices. The grid spectra were recorded at $T$ = 350 mK under an applied out-of-plane magnetic field of 0.65 T. (d) Layer-resolved superconducting gap spectra in a MnTe/Bi$_2$Te$_3$/Fe(Se,Te) heterostructure. (e) Anomalous Hall resistivity of the V$_5$Se$_8$/NbSe$_2$ heterostructure as a function of the magnetic fields taken at $T$ = 2 K with two different magnetic field angles ($\theta$ = 0° and 80°). (f) The anomalous Hall conductivity at $\mu_0 H$ = 9 T plotted against $\theta$. The yellow-colored area highlights a deviation from cos($\theta$). Panel (a) is reproduced with permission from Ref.[169]. Panels (b)-(c) are reproduced with permission from Ref.[170]. Panel (d) is reproduced with permission from Ref.[29]. Panel (e) is reproduced with permission from Ref.[17]. Panel (f) is reproduced with permission from Ref.[18].

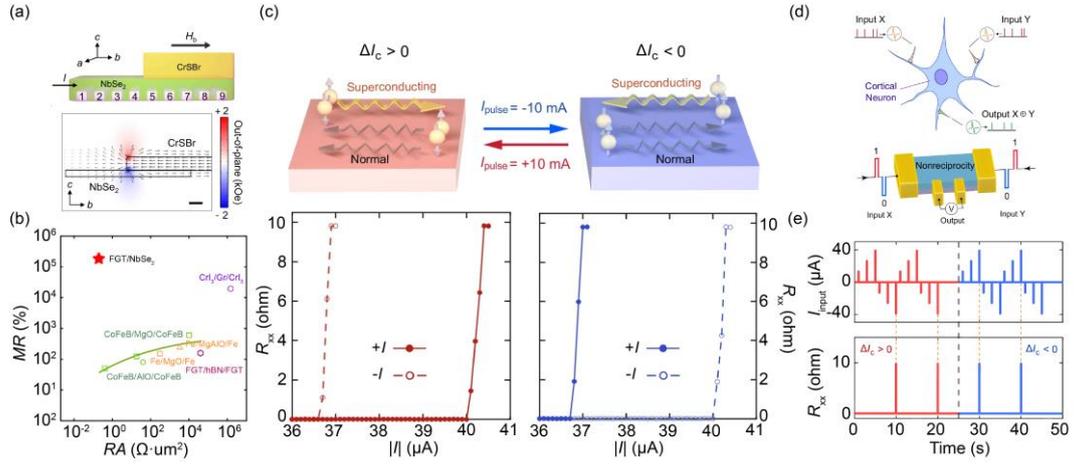

**Fig. 7.** Giant magnetoresistance and electrically switchable nonreciprocal superconducting transport in the superconductor/magnet heterostructure. (a) Device schematic of a NbSe$_2$/CrSBr heterostructure, with the CrSBr edge positioned between electrodes, and simulated out-of-plane stray-field distribution. (b) Comparison of magnetoresistance and resistance–area product with representative conventional and superconducting spintronic devices reported in the literature. (c) Schematics and experimental demonstration of electrically switchable nonreciprocal superconducting transport. (d) Conceptual analogy between biological cortical neurons and neural transistors, and spike responses to current pulses for opposite polarity states, illustrating the potential of nonreciprocal superconducting devices for neuromorphic computing. Panel (a) is reproduced with permission from Ref.[153]. Panels (b)-(e) are reproduced with permission from Ref.[15].

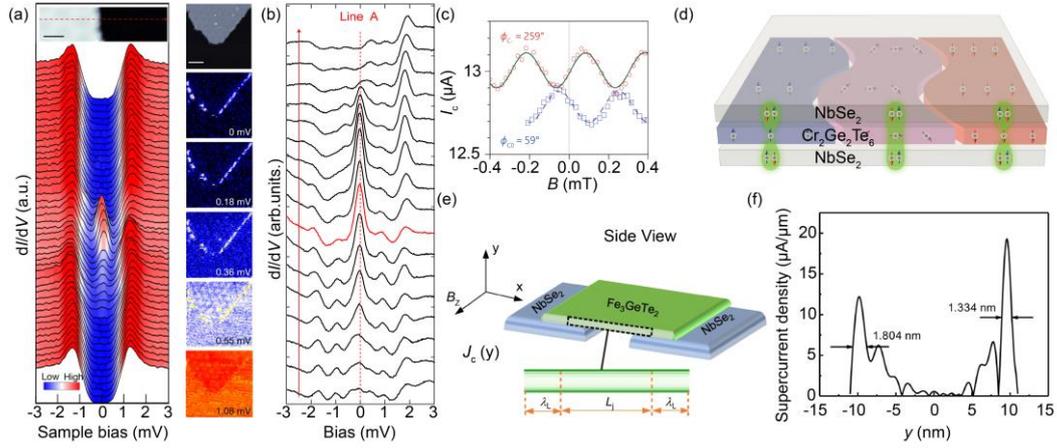

**Fig. 8.** (a) Differential conductance spectra (d$I$/d$V$) measured across the edge of a CrBr$_3$ island, with the corresponding STM topography shown above. (b) Spatially resolved d$I$/d$V$ spectra taken along a vortex core (Line A). (c) SQUID oscillations of a Cr$_2$Ge$_2$Te$_6$-based Josephson junction, where red circles and blue squares denote the critical currents $I_C^-$ and $I_C^0$ measured at $T = 0.9$ K. (d) Schematic illustration of magnetization-dependent Josephson coupling: perpendicular magnetization favors spin-conserving tunneling of Ising Cooper pairs and a π-phase junction, while in-plane magnetization requires spin-flip tunneling and yields a 0-phase junction, resulting in mixed 0–π segments in the presence of magnetic domains. (e) Schematic of the S/F/S junction with the magnetic field along the y-axis direction. (f) Distribution of supercurrent density along the y-axis, revealing two conductive edge channels. Panel (a) is reproduced with permission from Ref.[28]. Panel (b) is reproduced with permission from Ref.[29]. Panels (c)-(d) are reproduced with permission from Ref.[162]. Panels (e)-(f) are reproduced with permission from Ref.[165].

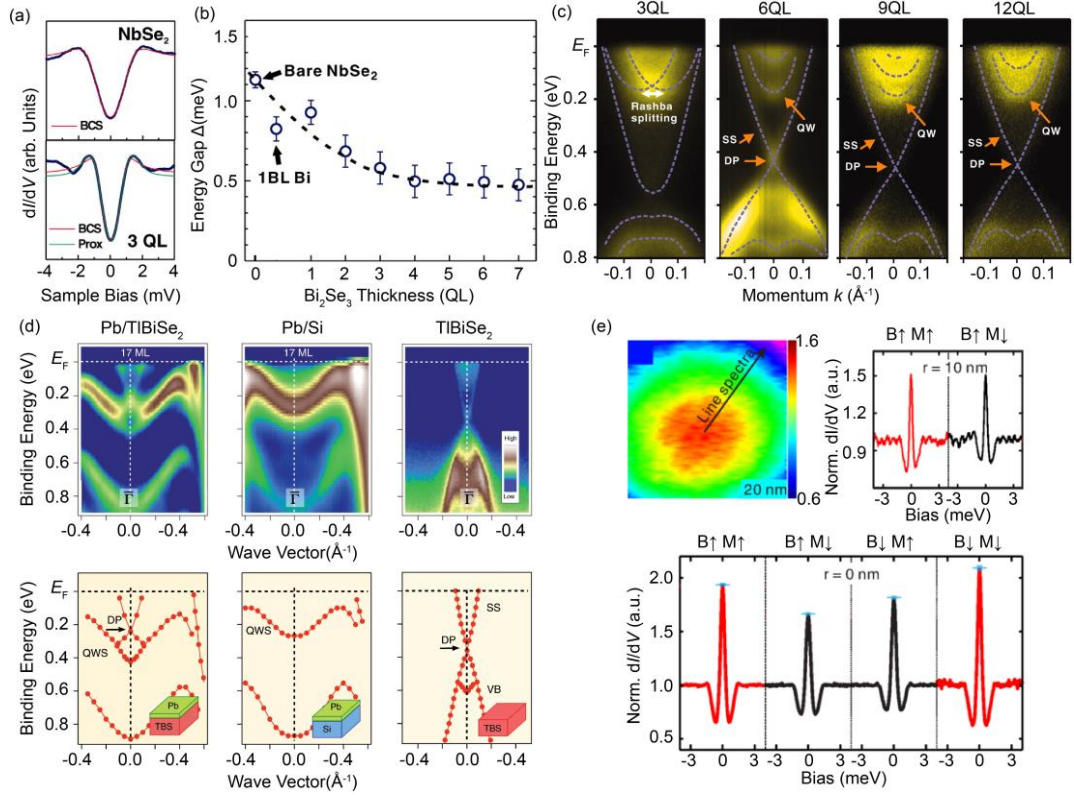

**Fig.9.** Proximity-induced topological superconductivity and Majorana signatures in superconductor/topological material heterostructures. (a) The STS spectra measured on pristine NbSe$_2$ and a 3QL-Bi$_2$Se$_3$ film. (b) Thickness dependence of the induced superconducting gap extracted from spectral fitting. (c) ARPES-measured band dispersions of Bi$_2$Se$_3$ thin films with varying thickness. (d) (Upper panels) ARPES-derived band structures near the Fermi level around th $\overline{\Gamma}$ point for 17ML-Pb/TlBiSe$_2$, 17ML-Pb/Si(111), and pristine TlBiSe$_2$, respectively. (Lower panels) Corresponding experimental dispersions obtained from MDC/EDC peak positions. (e) (Left) Zero bias d$I$/d$V$ mapping of a vortex at 0.1 T with the spin nonpolarized tip on the topological superconductor 5QL Bi$_2$Te$_3$/NbSe$_2$; (Lower) d$I$/d$V$ spectrum measured at the vortex center using a fully spin-polarized tip. (Right) d$I$/d$V$ spectrum taken ~10 nm away from the vortex center with a fully spin-polarized tip. Panels (a)-(c) are reproduced with permission from Ref.[185]. Panel (d) is reproduced with permission from Ref.[199]. Panel (e) is reproduced with permission from Ref.[198].

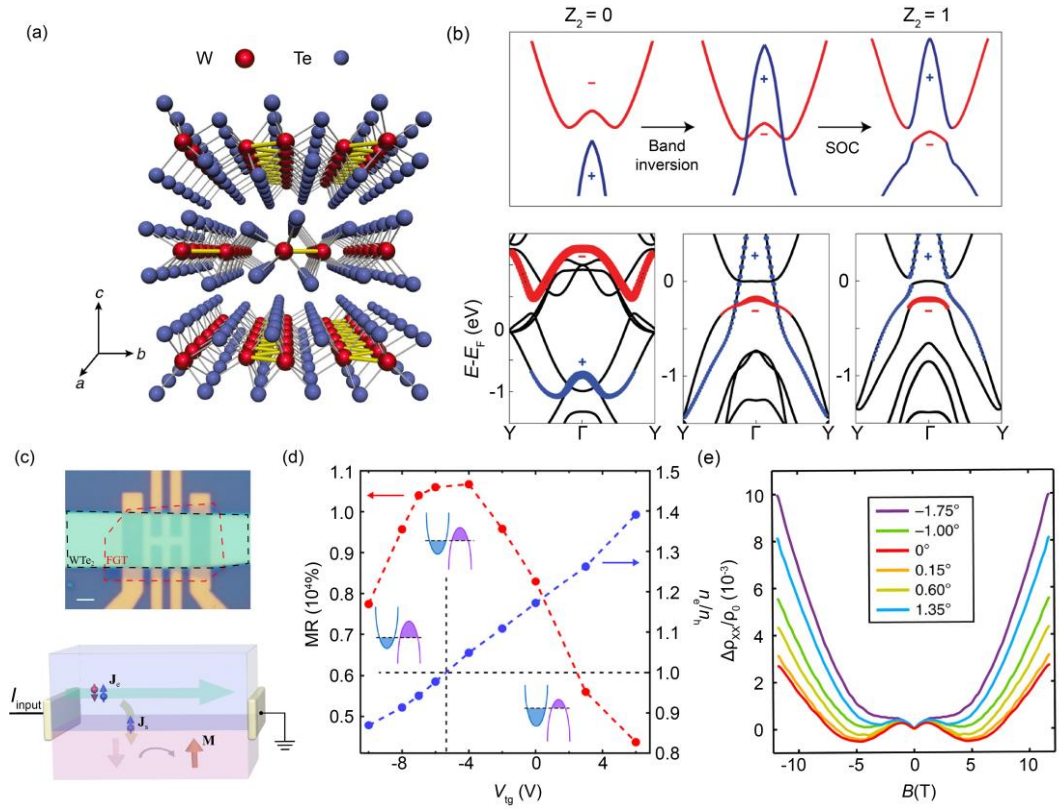

**Fig. 10.** Electronic structure and emergent phenomena in topological semimetal. (a) The crystal structure of $WTe_2$. (b) (Upper panel) Schematic diagram to show the bulk band evolution from a topologically trivial phase, to a non-trivial phase, and then to a bulk band opening due to SOC. (Lower panels) Calculated band structures for $WTe_2$ to show the evolution from 1T-$WTe_2$ to 1T′-$WTe_2$ without SOC and finally 1T′-$WTe_2$ with SOC. (c) (Upper panel) Optical image of a FGT/$WTe_2$ device. (Lower panel) Schematic of the current induced spin polarization in the $WTe_2$ flake. (d) Comparison of gate-voltage-dependent magnetoresistance (MR) and carrier density ratios. (e) Negative longitudinal MR in thin $WTe_2$. Panel (a) is reproduced with permission from Ref.[203]. Panel (b) is reproduced with permission from Ref.[181]. Panel (c) is reproduced with permission from Ref.[205]. Panels (d) is reproduced with permission from Ref.[204]. Panel (e) is reproduced with permission from Ref.[203].

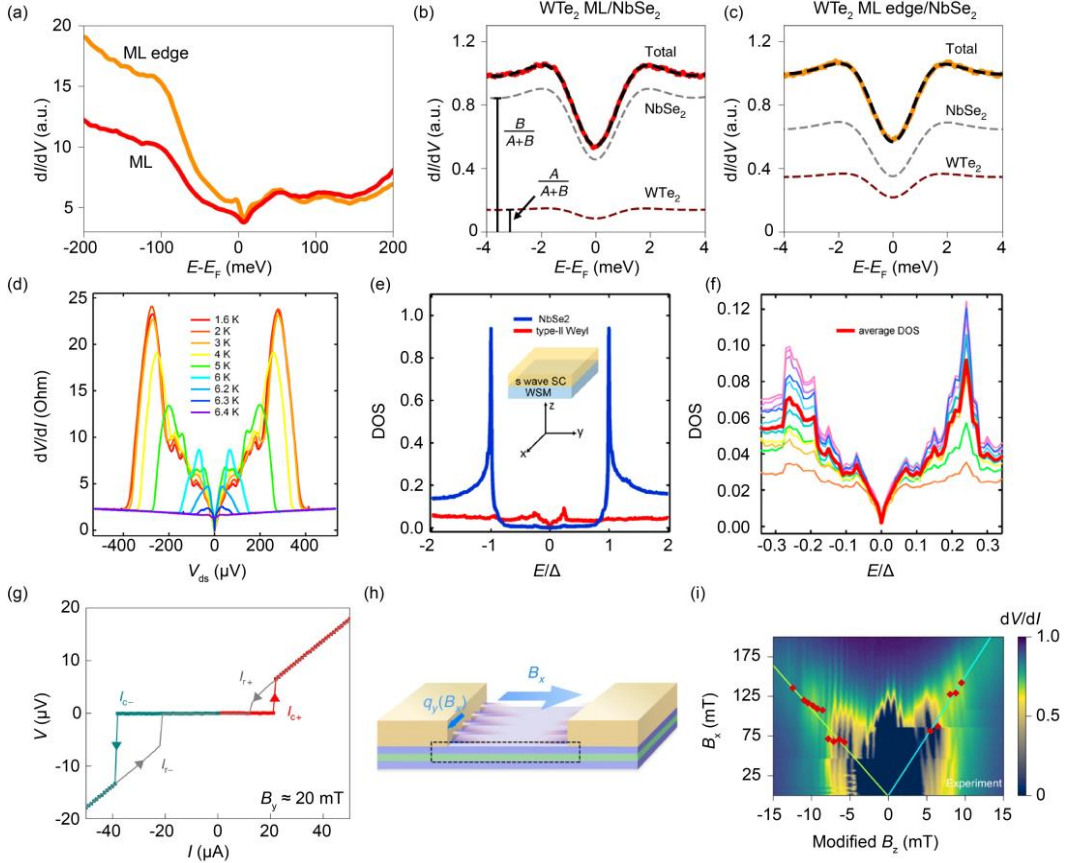

**Fig. 11.** Emergent topological and nonreciprocal superconducting phenomena in superconductor/topological semimetal heterostructure. (a) Representative $dI/dV$ spectra of monolayer $WTe_2$ away from the edge (bulk) and at the edge, showing enhanced local density of states (LDOS) due to the quantum spin Hall edge state. (b-c) Fitting of representative tunneling spectra for monolayer $WTe_2$/ $NbSe_2$ in the bulk (b) and at the edge (c). (d) Differential resistance spectra of $WTe_2$ measured at various temperatures, revealing subgap features inside the main superconducting peaks. (e-f) Theoretical calculations of density of states (DOS) in superconducting $WTe_2/NbSe_2$ heterostructure. (e) DOS spectra of type II Weyl semi-metal ($WTe_2$) and s-wave superconductor ($NbSe_2$) in the superconducting state. (f) Enlarged DOS spectra of $WTe_2$ in panel a (red line). Thin curves represent DOS of different layers in $WTe_2$. (g) $I$–$V$ curve of a $NiTe_2$/Nb JJ device (d = 350 nm) at 20 mK under in-plane field $B_y$ = 20 mT, demonstrating nonreciprocal critical currents. (h) Schematic of the junction under in-plane field along current, showing formation of a finite-momentum Cooper pair modulation in the $y$ direction. (i) Dependence of $dV/dI$ on in-plane ($B_x$) and out-of-plane ($B_z$) magnetic fields for the same JJ device with $d$ = 350 nm, showing the evolution of the interference pattern due to the applied in-plane magnetic field. Panels (a)-(c) are reproduced with permission from Ref.[188]. Panels (d)-(f) are reproduced with permission from Ref.[208]. Panels (g)-(i) are reproduced with permission from Ref.[23].

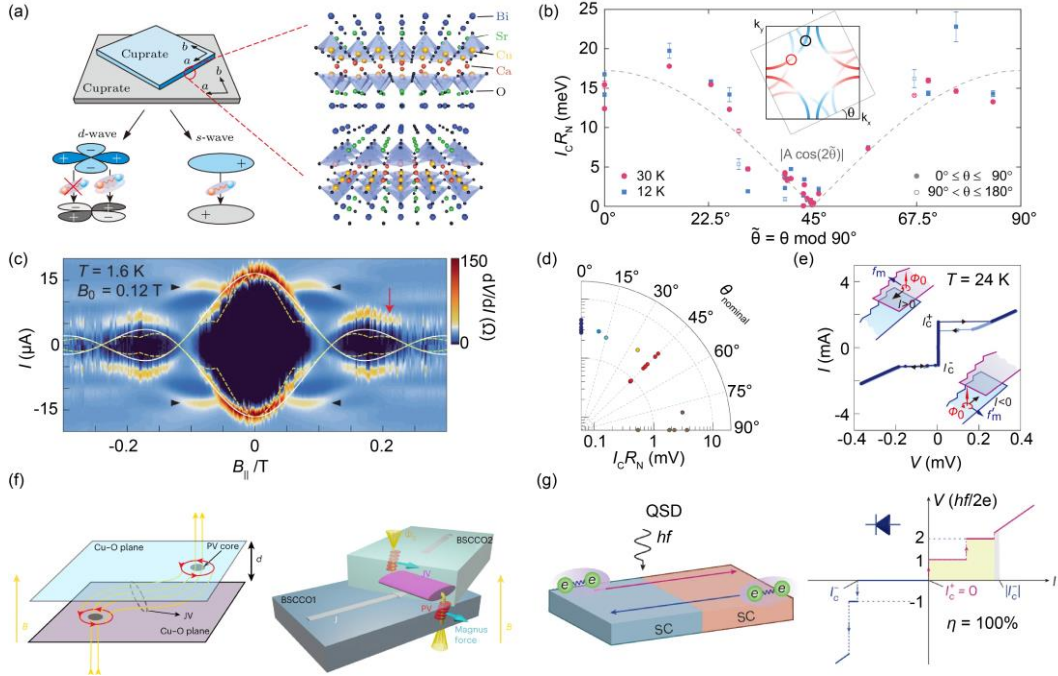

**Fig. 12.** Twist-controlled Josephson physics in cuprate superconductors. (a) Left: schematic of a twisted cuprate bicrystal illustrating Josephson tunneling for $s$-, $d$-, or $d + id$-wave pairing. Right: atomic structure of a 45°-twisted $Bi_2Sr_2CaCu_2O_8$ (Bi-2212) interface. (b) Angular dependence of $I_C R_N$ at 30 K and 12 K, showing nearly incoherent tunneling. (c) Fraunhofer diffraction pattern for a 44.8°-twisted Bi-2201 Josephson junction. (d) Josephson coupling strength as a function of twist angle in underdoped Bi-2212. (e) Experimental observation of Josephson diode effect in twisted cuprates. (f) Left: schematic of a twisted BSCCO junction with vortex configuration. Right: bias-current-induced changes in vortex configuration (Magnus force) generate nonreciprocal transport. (g) Schematic of a quantized superconducting diode (QSD) under microwave irradiation (left) and its I–V characteristic showing perfect diode efficiency (right). Panel (a) is reproduced with permission from Ref.[211]. Panel (b) is reproduced with permission from Ref.[66]. Panel (c) is reproduced with permission from Ref.[65]. Panel (d) is reproduced with permission from Ref.[63]. Panel (e) is reproduced with permission from Ref.[215]. Panel (f) is reproduced with permission from Ref.[212]. Panel (g) is reproduced with permission from Ref.[213].

# References

[1] Dvir T, Massee F, Attias L, Khodas M, Aprili M, Quay C H L and Steinberg H 2018 *Nat. Commun.* **9** 598

[2] Khestanova E, Birkbeck J, Zhu M, Cao Y, Yu G L, Ghazaryan D, Yin J, Berger H, Forró L, Taniguchi T, Watanabe K, Gorbachev R V, Mishchenko A, Geim A K and Grigorieva I V 2018 *Nano Lett.* **18** 2623

[3] Yan R, Khalsa G, Schaefer B T, Jarjour A, Rouvimov S, Nowack K C, Xing H G and Jena D 2019 *Appl. Phys. Express* **12** 023008

[4] Wu Y, He J, Liu J, Xing H, Mao Z and Liu Y 2019 *Nanotechnology* **30** 035702

[5] Saito Y, Nojima T and Iwasa Y 2018 *Nat. Commun.* **9** 778

[6] Gozar A, Logvenov G, Kourkoutis L F, Bollinger A T, Giannuzzi L A, Muller D A and Bozovic I 2008 *Nature* **455** 782

[7] Wang Q-Y, Li Z, Zhang W-H, Zhang Z-C, Zhang J-S, Li W, Ding H, Ou Y-B, Deng P, Chang K, Wen J, Song C-L, He K, Jia J-F, Ji S-H, Wang Y-Y, Wang L-L, Chen X, Ma X-C and Xue Q-K 2012 *Chin. Phys. Lett.* **29** 037402

[8] Zhou Z, Hou F, Huang X, Wang G, Fu Z, Liu W, Yuan G, Xi X, Xu J, Lin J and Gao L 2023 *Nature* **621** 499

[9] Yin S, Wang R, Wang L, Liu Y, Zhu C, Zhou Z, Ji Q, Gao L and Liu X 2025 *Nano Lett.* **25** 10603

[10] Zhang G, Wu H, Jin W, Yang L, Xiao B, Yu J, Zhang W and Chang H 2025 *Cell Rep. Phys. Sci.* **6** 102356

[11] Nam H, Chen H, Liu T, Kim J, Zhang C, Yong J, Lemberger T R, Kratz P A, Kirtley J R, Moler K, Adams P W, MacDonald A H and Shih C-K 2016 *Proceedings of the National Academy of Sciences* **113** 10513

[12] Wang C, Lian B, Guo X, Mao J, Zhang Z, Zhang D, Gu B-L, Xu Y and Duan W 2019 *Phys. Rev. Lett.* **123** 126402

[13] You J-Y, Gu B, Su G and Feng Y P 2021 *Phys. Rev. B* **103** 104503

[14] Zhang Z, Wu Z, Fang C, Zhang F-c, Hu J, Wang Y and Qin S 2024 *Nat. Commun.* **15** 7971

[15] Xiong J, Xie J, Cheng B, Dai Y, Cui X, Wang L, Liu Z, Zhou J, Wang N, Xu X, Chen X, Cheong S-W, Liang S-J and Miao F 2024 *Nat. Commun.* **15** 4953

[16] de la Barrera S C, Sinko M R, Gopalan D P, Sivadas N, Seyler K L, Watanabe K, Taniguchi T, Tsen A W, Xu X, Xiao D and Hunt B M 2018 *Nat. Commun.* **9** 1427

[17] Matsuoka H, Barnes S E, Ieda J i, Maekawa S, Bahramy M S, Saika B K, Takeda Y, Wadati H, Wang Y, Yoshida S, Ishizaka K, Iwasa Y and Nakano M 2021 *Nano Lett.* **21** 1807

[18] Matsuoka H, Habe T, Iwasa Y, Koshino M and Nakano M 2022 *Nat. Commun.* **13** 5129

[19] Kontos T, Aprili M, Lesueur J and Grison X 2001 *Phys. Rev. Lett.* **86** 304

[20] Chen A Q, Park M J, Gill S T, Xiao Y, Reig-i-Plessis D, MacDougall G J, Gilbert M J and Mason N 2018 *Nat. Commun.* **9** 3478

[21] Jiang D, Yuan T, Wu Y, Wei X, Mu G, An Z and Li W 2020 *ACS Appl. Mater. Interfaces* **12** 49252

[22] Cai R, Yao Y, Lv P, Ma Y, Xing W, Li B, Ji Y, Zhou H, Shen C, Jia S, Xie X C, Žutić I, Sun Q-F and Han W 2021 *Nat. Commun.* **12** 6725

[23] Pal B, Chakraborty A, Sivakumar P K, Davydova M, Gopi A K, Pandeya A K, Krieger J A,

Zhang Y, Date M, Ju S, Yuan N, Schröter N B M, Fu L and Parkin S S P 2022 *Nat. Phys.* **18** 1228

[24] Braunecker B and Simon P 2013 *Phys. Rev. Lett.* **111** 147202
[25] Nadj-Perge S, Drozdov I K, Bernevig B A and Yazdani A 2013 *Phys. Rev. B* **88** 020407
[26] Pientka F, Glazman L I and von Oppen F 2013 *Phys. Rev. B* **88** 155420
[27] Nakosai S, Tanaka Y and Nagaosa N 2013 *Phys. Rev. B* **88** 180503
[28] Kezilebieke S, Huda M N, Vaňo V, Aapro M, Ganguli S C, Silveira O J, Głodzik S, Foster A S, Ojanen T and Liljeroth P 2020 *Nature* **588** 424
[29] Ding S, Chen C, Cao Z, Wang D, Pan Y, Tao R, Zhao D, Hu Y, Jiang T, Yan Y, Shi Z, Wan X, Feng D and Zhang T 2022 *Sci. Adv.* **8** eabq4578
[30] Bardeen J, Cooper L N and Schrieffer J R 1957 *Physical Review* **106** 162
[31] Bardeen J, Cooper L N and Schrieffer J R 1957 *Physical Review* **108** 1175
[32] Abrahams E, Balatsky A, Scalapino D J and Schrieffer J R 1995 *Phys. Rev. B* **52** 1271
[33] Bergeret F S, Volkov A F and Efetov K B 2001 *Phys. Rev. B* **64** 134506
[34] Buzdin A I 2005 *Rev. Mod. Phys.* **77** 935
[35] Asano Y, Sawa Y, Tanaka Y and Golubov A A 2007 *Phys. Rev. B* **76** 224525
[36] Halterman K, Valls O T and Barsic P H 2008 *Phys. Rev. B* **77** 174511
[37] Huxley A D 2015 *Physica C* **514** 368
[38] Ran S, Eckberg C, Ding Q-P, Furukawa Y, Metz T, Saha S R, Liu I L, Zic M, Kim H, Paglione J and Butch N P 2019 *Science* **365** 684
[39] Bergeret F S, Volkov A F and Efetov K B 2001 *Phys. Rev. Lett.* **86** 4096
[40] Houzet M and Buzdin A I 2007 *Phys. Rev. B* **76** 060504
[41] Linder J and Robinson J W A 2015 *Nat. Phys.* **11** 307
[42] Qi X-L and Zhang S-C 2011 *Rev. Mod. Phys.* **83** 1057
[43] Sato M and Ando Y 2017 *Rep. Prog. Phys.* **80** 076501
[44] Hu J, Yu F, Luo A, Pan X-H, Zou J, Liu X and Xu G 2024 *Phys. Rev. Lett.* **132** 036601
[45] Fu L and Kane C L 2008 *Phys. Rev. Lett.* **100** 096407
[46] Choi Y-B, Xie Y, Chen C-Z, Park J, Song S-B, Yoon J, Kim B J, Taniguchi T, Watanabe K, Kim J, Fong K C, Ali M N, Law K T and Lee G-H 2020 *Nat. Mater.* **19** 974
[47] Novoselov K S, Mishchenko A, Carvalho A and Castro Neto A H 2016 *Science* **353** aac9439
[48] Gong C and Zhang X 2019 *Science* **363** eaav4450
[49] Liu Y, Huang Y and Duan X 2019 *Nature* **567** 323
[50] Yi H, Zhao Y-F, Chan Y-T, Cai J, Mei R, Wu X, Yan Z-J, Zhou L-J, Zhang R, Wang Z, Paolini S, Xiao R, Wang K, Richardella A R, Singleton J, Winter L E, Prokscha T, Salman Z, Suter A, Balakrishnan P P, Grutter A J, Chan M H W, Samarth N, Xu X, Wu W, Liu C-X and Chang C-Z 2024 *Science* **383** 634
[51] Liu X, Shan J, Cao T, Zhu L, Ma J, Wang G, Shi Z, Yang Q, Ma M, Liu Z, Yan S, Wang L, Dai Y, Xiong J, Chen F, Wang B, Pan C, Wang Z, Cheng B, He Y, Luo X, Lin J, Liang S-J and Miao F 2024 *Nat. Mater.* **23** 1363
[52] Ye J T, Inoue S, Kobayashi K, Kasahara Y, Yuan H T, Shimotani H and Iwasa Y 2010 *Nat. Mater.* **9** 125
[53] Lu J M, Zheliuk O, Leermakers I, Yuan N F Q, Zeitler U, Law K T and Ye J T 2015 *Science* **350** 1353
[54] Saito Y, Kasahara Y, Ye J, Iwasa Y and Nojima T 2015 *Science* **350** 409

[55] Xi X, Berger H, Forró L, Shan J and Mak K F 2016 *Phys. Rev. Lett.* **117** 106801

[56] Costanzo D, Jo S, Berger H and Morpurgo A F 2016 *Nat. Nanotechnol.* **11** 339

[57] Fu Y, Liu E, Yuan H, Tang P, Lian B, Xu G, Zeng J, Chen Z, Wang Y, Zhou W, Xu K, Gao A, Pan C, Wang M, Wang B, Zhang S-C, Cui Y, Hwang H Y and Miao F 2017 *npj Quantum Mater.* **2** 52

[58] Huang B, Clark G, Klein D R, MacNeill D, Navarro-Moratalla E, Seyler K L, Wilson N, McGuire M A, Cobden D H, Xiao D, Yao W, Jarillo-Herrero P and Xu X 2018 *Nat. Nanotechnol.* **13** 544

[59] Wang Z, Zhang T, Ding M, Dong B, Li Y, Chen M, Li X, Huang J, Wang H, Zhao X, Li Y, Li D, Jia C, Sun L, Guo H, Ye Y, Sun D, Chen Y, Yang T, Zhang J, Ono S, Han Z and Zhang Z 2018 *Nat. Nanotechnol.* **13** 554

[60] Zeng J, Liu E, Fu Y, Chen Z, Pan C, Wang C, Wang M, Wang Y, Xu K, Cai S, Yan X, Wang Y, Liu X, Wang P, Liang S-J, Cui Y, Hwang H Y, Yuan H and Miao F 2018 *Nano Lett.* **18** 1410

[61] Xie J, Xiong J-L, Cheng B, Liang S-J and Miao F 2025 *Chin. Phys. Lett.* **42** 040202

[62] Cao Y, Fatemi V, Fang S, Watanabe K, Taniguchi T, Kaxiras E and Jarillo-Herrero P 2018 *Nature* **556** 43

[63] Zhu Y, Liao M, Zhang Q, Xie H-Y, Meng F, Liu Y, Bai Z, Ji S, Zhang J, Jiang K, Zhong R, Schneeloch J, Gu G, Gu L, Ma X, Zhang D and Xue Q-K 2021 *Phys. Rev. X* **11** 031011

[64] Li Q, Cheng B, Chen M, Xie B, Xie Y, Wang P, Chen F, Liu Z, Watanabe K, Taniguchi T, Liang S-J, Wang D, Wang C, Wang Q-H, Liu J and Miao F 2022 *Nature* **609** 479

[65] Wang H, Zhu Y, Bai Z, Wang Z, Hu S, Xie H-Y, Hu X, Cui J, Huang M, Chen J, Ding Y, Zhao L, Li X, Zhang Q, Gu L, Zhou X J, Zhu J, Zhang D and Xue Q-K 2023 *Nat. Commun.* **14** 5201

[66] Zhao S Y F, Cui X, Volkov P A, Yoo H, Lee S, Gardener J A, Akey A J, Engelke R, Ronen Y, Zhong R, Gu G, Plugge S, Tummuru T, Kim M, Franz M, Pixley J H, Poccia N and Kim P 2023 *Science* **382** 1422

[67] Chen M, Xie Y, Cheng B, Yang Z, Li X-Z, Chen F, Li Q, Xie J, Watanabe K, Taniguchi T, He W-Y, Wu M, Liang S-J and Miao F 2024 *Nat. Nanotechnol.* **19** 962

[68] Wang Y, Wang C, Liang S-J, Ma Z, Xu K, Liu X, Zhang L, Admasu A S, Cheong S-W, Wang L, Chen M, Liu Z, Cheng B, Ji W and Miao F 2020 *Adv. Mater.* **32** 2004533

[69] Miao F, Liang S-J and Cheng B 2021 *npj Quantum Mater.* **6** 59

[70] Ma Z, Yan S, Chen F, Dai Y, Liu Z, Xu K, Xu T, Tong Z, Chen M, Wang L, Wang P, Sun L, Cheng B, Liang S-J and Miao F 2024 *Chin. Phys. Lett.* **41**

[71] Liu Z, Shi J, Cao J, Ma Z, Yang Z, Cui Y, Wang L, Dai Y, Chen M, Wang P, Xie Y, Chen F, Shi Y, Xiao C, Yang S A, Cheng B, Liang S-J and Miao F 2025 *Adv. Funct. Mater.* **35** 2416204

[72] Wang C-Y, Wang C, Meng F, Wang P, Wang S, Liang S-J and Miao F 2020 *Adv. Electron. Mater.* **6** 1901107

[73] Liang S-J, Cheng B, Cui X and Miao F 2020 *Adv. Mater.* **32** 1903800

[74] Liang S-J, Li Y, Cheng B and Miao F 2022 *Small Struct.* **3** 2200064

[75] Pan X, Li Y, Cheng B, Liang S-J and Miao F 2023 *Sci. China Phys., Mech. Astron.* **66** 117504

[76] Manzeli S, Ovchinnikov D, Pasquier D, Yazyev O V and Kis A 2017 *Nat. Rev. Mater.* **2**

17033

[77] Xi X, Wang Z, Zhao W, Park J-H, Law K T, Berger H, Forró L, Shan J and Mak K F 2016 *Nat. Phys.* **12** 139

[78] Xi X, Zhao L, Wang Z, Berger H, Forró L, Shan J and Mak K F 2015 *Nat. Nanotechnol.* **10** 765

[79] Wickramaratne D, Khmelevskyi S, Agterberg D F and Mazin I I 2020 *Phys. Rev. X* **10** 041003

[80] Wan P, Zheliuk O, Yuan N F Q, Peng X, Zhang L, Liang M, Zeitler U, Wiedmann S, Hussey N E, Palstra T T M and Ye J 2023 *Nature* **619** 46

[81] Cho C-w, Lyu J, An L, Han T, Lo K T, Ng C Y, Hu J, Gao Y, Li G, Huang M, Wang N, Schmalian J and Lortz R 2022 *Phys. Rev. Lett.* **129** 087002

[82] Sticlet D and Morari C 2019 *Phys. Rev. B* **100** 075420

[83] Shaffer D, Kang J, Burnell F J and Fernandes R M 2020 *Phys. Rev. B* **101** 224503

[84] Hamill A, Heischmidt B, Sohn E, Shaffer D, Tsai K-T, Zhang X, Xi X, Suslov A, Berger H, Forró L, Burnell F J, Shan J, Mak K F, Fernandes R M, Wang K and Pribiag V S 2021 *Nat. Phys.* **17** 949

[85] Ahadi K, Galletti L, Li Y, Salmani-Rezaie S, Wu W and Stemmer S 2019 *Sci. Adv.* **5** eaaw0120

[86] Chen W, Liu Z, Huo Z, Dong G, Cai J and Duan D 2024 *Chin. Phys. Lett.* **41** 117403

[87] Liu P, Lei B, Chen X, Wang L and Wang X 2022 *Nat. Rev. Phys.* **4** 336

[88] Nakagawa Y, Kasahara Y, Nomoto T, Arita R, Nojima T and Iwasa Y 2021 *Science* **372** 190

[89] Ye J T, Zhang Y J, Akashi R, Bahramy M S, Arita R and Iwasa Y 2012 *Science* **338** 1193

[90] Cui Y, Liu Z, Liu Q, Xiong J, Xie Y, Dai Y, Zhou J, Wang L, Fang H, Liu H, Liang S-J, Cheng B and Miao F 2025 *Phys. Rev. Lett.* **135** 076501

[91] Bøttcher C G L, Nichele F, Kjaergaard M, Suominen H J, Shabani J, Palmstrøm C J and Marcus C M 2018 *Nat. Phys.* **14** 1138

[92] Yang C, Liu Y, Wang Y, Feng L, He Q, Sun J, Tang Y, Wu C, Xiong J, Zhang W, Lin X, Yao H, Liu H, Fernandes G, Xu J, Valles J M, Wang J and Li Y 2019 *Science* **366** 1505

[93] Kapitulnik A, Kivelson S A and Spivak B 2019 *Rev. Mod. Phys.* **91** 011002

[94] Yang C, Liu H, Liu Y, Wang J, Qiu D, Wang S, Wang Y, He Q, Li X, Li P, Tang Y, Wang J, Xie X C, Valles J M, Xiong J and Li Y 2022 *Nature* **601** 205

[95] Li Y, Liu H, Ji H, Ji C, Qi S, Jiao X, Dong W, Sun Y, Zhang W, Cui Z, Pan M, Samarth N, Wang L, Xie X C, Xue Q-K, Liu Y and Wang J 2024 *Phys. Rev. Lett.* **132** 226003

[96] Tamir I, Benyamini A, Telford E J, Gorniaczyk F, Doron A, Levinson T, Wang D, Gay F, Sacépé B, Hone J, Watanabe K, Taniguchi T, Dean C R, Pasupathy A N and Shahar D 2019 *Sci. Adv.* **5** eaau3826

[97] Xing Y, Yang P, Ge J, Yan J, Luo J, Ji H, Yang Z, Li Y, Wang Z, Liu Y, Yang F, Qiu P, Xi C, Tian M, Liu Y, Lin X and Wang J 2021 *Nano Lett.* **21** 7486

[98] Yamada K, Lee C H, Kurahashi K, Wada J, Wakimoto S, Ueki S, Kimura H, Endoh Y, Hosoya S, Shirane G, Birgeneau R J, Greven M, Kastner M A and Kim Y J 1998 *Phys. Rev. B* **57** 6165

[99] Hoffman J E, Hudson E W, Lang K M, Madhavan V, Eisaki H, Uchida S and Davis J C 2002 *Science* **295** 466

[100] Howald C, Eisaki H, Kaneko N and Kapitulnik A 2003 *Proceedings of the National*

        *Academy of Sciences of the United States of America* **100** 9705

[101] Wang Y, Li L and Ong N P 2006 *Phys. Rev. B* **73** 024510

[102] Doiron-Leyraud N, Proust C, LeBoeuf D, Levallois J, Bonnemaison J-B, Liang R, Bonn D A, Hardy W N and Taillefer L 2007 *Nature* **447** 565

[103] Motoyama E M, Yu G, Vishik I M, Vajk O P, Mang P K and Greven M 2007 *Nature* **445** 186

[104] Wu T, Mayaffre H, Krämer S, Horvatić M, Berthier C, Hardy W N, Liang R, Bonn D A and Julien M-H 2011 *Nature* **477** 191

[105] Fradkin E, Kivelson S A and Tranquada J M 2015 *Rev. Mod. Phys.* **87** 457

[106] Caprara S, Di Castro C, Seibold G and Grilli M 2017 *Phys. Rev. B* **95** 224511

[107] Bednorz J G and Müller K A 1986 *Zeitschrift für Physik B Condensed Matter* **64** 189

[108] Wu M K, Ashburn J R, Torng C J, Hor P H, Meng R L, Gao L, Huang Z J, Wang Y Q and Chu C W 1987 *Phys. Rev. Lett.* **58** 908

[109] Maeda H, Tanaka Y, Fukutomi M and Asano T 1988 *Jpn. J. Appl. Phys.* **27** L209

[110] Yu Y, Ma L, Cai P, Zhong R, Ye C, Shen J, Gu G D, Chen X H and Zhang Y 2019 *Nature* **575** 156

[111] McElroy K, Simmonds R W, Hoffman J E, Lee D H, Orenstein J, Eisaki H, Uchida S and Davis J C 2003 *Nature* **422** 592

[112] McElroy K, Lee J, Slezak J A, Lee D H, Eisaki H, Uchida S and Davis J C 2005 *Science* **309** 1048

[113] Jiang D, Hu T, You L, Li Q, Li A, Wang H, Mu G, Chen Z, Zhang H, Yu G, Zhu J, Sun Q, Lin C, Xiao H, Xie X and Jiang M 2014 *Nat. Commun.* **5** 5708

[114] Bollinger A T and Božović I 2016 *Supercond. Sci. Technol.* **29** 103001

[115] Liao M, Zhu Y, Zhang J, Zhong R, Schneeloch J, Gu G, Jiang K, Zhang D, Ma X and Xue Q-K 2018 *Nano Lett.* **18** 5660

[116] Qi S, Ge J, Ji C, Ai Y, Ma G, Wang Z, Cui Z, Liu Y, Wang Z and Wang J 2025 *Nat. Commun.* **16** 531

[117] Orenstein J and Millis A J 2000 *Science* **288** 468

[118] Wan S, Li H, Choubey P, Gu Q, Li H, Yang H, Eremin I M, Gu G and Wen H-H 2021 *Proceedings of the National Academy of Sciences* **118** e2115317118

[119] Zareapour P, Hayat A, Zhao S Y F, Kreshchuk M, Jain A, Kwok D C, Lee N, Cheong S-W, Xu Z, Yang A, Gu G D, Jia S, Cava R J and Burch K S 2012 *Nat. Commun.* **3** 1056

[120] Wang E, Ding H, Fedorov A V, Yao W, Li Z, Lv Y-F, Zhao K, Zhang L-G, Xu Z, Schneeloch J, Zhong R, Ji S-H, Wang L, He K, Ma X, Gu G, Yao H, Xue Q-K, Chen X and Zhou S 2013 *Nat. Phys.* **9** 621

[121] Zahid Hasan M, Xu S-Y and Neupane M. in *Topological Insulators*  55 (2015).

[122] Kamihara Y, Watanabe T, Hirano M and Hosono H 2008 *Journal of the American Chemical Society* **130** 3296

[123] Chen X H, Wu T, Wu G, Liu R H, Chen H and Fang D F 2008 *Nature* **453** 761

[124] Chen G-F, Li Z, Wu D, Dong J, Li G, Hu W-Z, Zheng P, Luo J-L and Wang N-L 2008 *Chin. Phys. Lett.* **25** 2235

[125] Rotter M, Tegel M and Johrendt D 2008 *Phys. Rev. Lett.* **101** 107006

[126] Tapp J H, Tang Z, Lv B, Sasmal K, Lorenz B, Chu P C W and Guloy A M 2008 *Phys. Rev. B* **78** 060505

[127] Kong L, Cao L, Zhu S, Papaj M, Dai G, Li G, Fan P, Liu W, Yang F, Wang X, Du S, Jin C, Fu L, Gao H-J and Ding H 2021 *Nat. Commun.* **12** 4146

[128] Wang X C, Liu Q Q, Lv Y X, Gao W B, Yang L X, Yu R C, Li F Y and Jin C Q 2008 *Solid State Commun.* **148** 538

[129] Hsu F-C, Luo J-Y, Yeh K-W, Chen T-K, Huang T-W, Wu P M, Lee Y-C, Huang Y-L, Chu Y-Y, Yan D-C and Wu M-K 2008 *Proceedings of the National Academy of Sciences* **105** 14262

[130] Zhang P, Yaji K, Hashimoto T, Ota Y, Kondo T, Okazaki K, Wang Z, Wen J, Gu G D, Ding H and Shin S 2018 *Science* **360** 182

[131] Shibauchi T, Carrington A and Matsuda Y 2014 *Annu. Rev. Condens. Matter Phys.* **5** 113

[132] Coldea A I 2021 *Front. Phys.* **Volume 8 - 2020**

[133] Fernandes R M, Coldea A I, Ding H, Fisher I R, Hirschfeld P J and Kotliar G 2022 *Nature* **601** 35

[134] Ma Y-X, Hao M-N, Li Q, Ma K, Li H-D, Zhang D, Sun R-J, Jin S-F, and Zhao C-C 2025 *Chin. Phys. B* **34** 67402

[135] Tan S, Zhang Y, Xia M, Ye Z, Chen F, Xie X, Peng R, Xu D, Fan Q, Xu H, Jiang J, Zhang T, Lai X, Xiang T, Hu J, Xie B and Feng D 2013 *Nat. Mater.* **12** 634

[136] He S, He J, Zhang W, Zhao L, Liu D, Liu X, Mou D, Ou Y-B, Wang Q-Y, Li Z, Wang L, Peng Y, Liu Y, Chen C, Yu L, Liu G, Dong X, Zhang J, Chen C, Xu Z, Chen X, Ma X, Xue Q and Zhou X J 2013 *Nat. Mater.* **12** 605

[137] Wang Z, Zhang P, Xu G, Zeng L K, Miao H, Xu X, Qian T, Weng H, Richard P, Fedorov A V, Ding H, Dai X and Fang Z 2015 *Phys. Rev. B* **92** 115119

[138] Roppongi M, Cai Y, Ogawa K, Liu S, Zhao G, Oudah M, Fujii T, Imamura K, Fang S, Ishihara K, Hashimoto K, Matsuura K, Mizukami Y, Pula M, Young C, Marković I, Bonn D A, Watanabe T, Yamashita A, Mizuguchi Y, Luke G M, Kojima K M, Uemura Y J and Shibauchi T 2025 *Nat. Commun.* **16** 6573

[139] McLaughlin N J, Wang H, Huang M, Lee-Wong E, Hu L, Lu H, Yan G Q, Gu G, Wu C, You Y-Z and Du C R 2021 *Nano Lett.* **21** 7277

[140] Farhang C, Zaki N, Wang J, Gu G, Johnson P D and Xia J 2023 *Phys. Rev. Lett.* **130** 046702

[141] Wang D, Kong L, Fan P, Chen H, Zhu S, Liu W, Cao L, Sun Y, Du S, Schneeloch J, Zhong R, Gu G, Fu L, Ding H and Gao H-J 2018 *Science* **362** 333

[142] Chen C, Jiang K, Zhang Y, Liu C, Liu Y, Wang Z and Wang J 2020 *Nat. Phys.* **16** 536

[143] Gong C, Li L, Li Z, Ji H, Stern A, Xia Y, Cao T, Bao W, Wang C, Wang Y, Qiu Z Q, Cava R J, Louie S G, Xia J and Zhang X 2017 *Nature* **546** 265

[144] Huang B, Clark G, Navarro-Moratalla E, Klein D R, Cheng R, Seyler K L, Zhong D, Schmidgall E, McGuire M A, Cobden D H, Yao W, Xiao D, Jarillo-Herrero P and Xu X 2017 *Nature* **546** 270

[145] Deng Y, Yu Y, Song Y, Zhang J, Wang N Z, Sun Z, Yi Y, Wu Y Z, Wu S, Zhu J, Wang J, Chen X H and Zhang Y 2018 *Nature* **563** 94

[146] Weber D, Trout A H, McComb D W and Goldberger J E 2019 *Nano Lett.* **19** 5031

[147] Wang N, Tang H, Shi M, Zhang H, Zhuo W, Liu D, Meng F, Ma L, Ying J, Zou L, Sun Z and Chen X 2019 *Journal of the American Chemical Society* **141** 17166

[148] Liu X-W, Xiong J-L, Wang L-Z, Liang S-J, Cheng B and Miao F 2022 *Acta Physica Sinica* **71** 127503

[149] Iturriaga H, Martinez L M, Mai T T, Biacchi A J, Augustin M, Hight Walker A R, Noufal M, Sreenivasan S T, Liu Y, Santos E J G, Petrovic C and Singamaneni S R 2023 *npj 2D Mater. Appl.* **7** 56

[150] Xiong J, Jiang J, Cui Y, Gao H, Zhou J, Liu Z, Zhang K, Cheng S, Wu K, Cheong S-W, Chang K, Liu Z, Yang H, Liang S-J, Cheng B and Miao F 2025 *Phys. Rev. Lett.* **135** 206701

[151] Bobkov G A, Bobkov A M and Bobkova I V 2024 *Phys. Rev. B* **110** 104506

[152] Cai R, Žutić I and Han W 2023 *Adv. Quantum Technol.* **6** 2200080

[153] Jo J, Peisen Y, Yang H, Mañas-Valero S, Baldoví J J, Lu Y, Coronado E, Casanova F, Bergeret F S, Gobbi M and Hueso L E 2023 *Nat. Commun.* **14** 7253

[154] Yun J, Son S, Shin J, Park G, Zhang K, Shin Y J, Park J-G and Kim D 2023 *Phys. Rev. Res.* **5** L022064

[155] Zeng X, Ye G, Yang F, Ye Q, Zhang L, Ma B, Liu Y, Xie M, Han G, Hao Y, Luo J, Lu X, Liu Y and Wang X 2023 *ACS Appl. Mater. Interfaces* **15** 57397

[156] Nadeem M, Fuhrer M S and Wang X 2023 *Nat. Rev. Phys.* **5** 558

[157] Robinson J W A, Witt J D S and Blamire M G 2010 *Science* **329** 59

[158] Anwar M S, Czeschka F, Hesselberth M, Porcu M and Aarts J 2010 *Phys. Rev. B* **82** 100501

[159] Kezilebieke S, Vaňo V, Huda M N, Aapro M, Ganguli S C, Liljeroth P and Lado J L 2022 *Nano Lett.* **22** 328

[160] Ma Q, Ai L, Zhang Y and Xiu F 2025 *Adv. Mater.* **n/a** e07866

[161] Bakurskiy S V, Klenov N V, Soloviev I I, Kupriyanov M Y and Golubov A A 2013 *Phys. Rev. B* **88** 144519

[162] Idzuchi H, Pientka F, Huang K F, Harada K, Gül Ö, Shin Y J, Nguyen L T, Jo N H, Shindo D, Cava R J, Canfield P C and Kim P 2021 *Nat. Commun.* **12** 5332

[163] Ai L, Zhang E, Yang J, Xie X, Yang Y, Jia Z, Zhang Y, Liu S, Li Z, Leng P, Cao X, Sun X, Zhang T, Kou X, Han Z, Xiu F and Dong S 2021 *Nat. Commun.* **12** 6580

[164] Keizer R S, Goennenwein S T B, Klapwijk T M, Miao G, Xiao G and Gupta A 2006 *Nature* **439** 825

[165] Hu G, Wang C, Wang S, Zhang Y, Feng Y, Wang Z, Niu Q, Zhang Z and Xiang B 2023 *Nat. Commun.* **14** 1779

[166] Wakatsuki R, Saito Y, Hoshino S, Itahashi Y M, Ideue T, Ezawa M, Iwasa Y and Nagaosa N 2017 *Sci. Adv.* **3** e1602390

[167] Tokura Y and Nagaosa N 2018 *Nat. Commun.* **9** 3740

[168] Daido A, Ikeda Y and Yanase Y 2022 *Phys. Rev. Lett.* **128** 037001

[169] Hu G, Wang C, Lu J, Zhu Y, Xi C, Ma X, Yang Y, Zhang Y, Wang S, Gu M, Zhang J, Lu Y, Cui P, Chen G, Zhu W, Xiang B and Zhang Z 2025 *ACS Nano* **19** 5709

[170] Kezilebieke S, Silveira O J, Huda M N, Vaňo V, Aapro M, Ganguli S C, Lahtinen J, Mansell R, van Dijken S, Foster A S and Liljeroth P 2021 *Adv. Mater.* **33** 2006850

[171] Che B-Y, Hu G-J, Zhu C, Guo H, Lv S-H, Liu X-Y, Wu K, Zhao Z, Pan L-L, Zhu K, Qi Q, Han Y-C, Lin X, Li Z-A, Shen C-M, Bao L-H, Liu Z, Zhou J-D, Yang H-T and Gao H-J 2024 *Chin. Phys. B* **33** 27502

[172] Li Y, Yin R, Li M, Gong J, Chen Z, Zhang J, Yan Y-J and Feng D-L 2024 *Nat. Commun.* **15** 10121

[173] Menezes R M, Neto J F S, Silva C C d S and Milošević M V 2019 *Phys. Rev. B* **100** 014431

[174] Andriyakhina E S and Burmistrov I S 2021 *Phys. Rev. B* **103** 174519

[175] Sun Y, Birch M T, Finizio S, Powalla L, Satheesh S, Priessnitz T, Göring E, Knöckl E, Kastl C, Holleitner A, Kern K, Weigand M, Wintz S and Burghard M 2025 *Adv. Mater.* **37** 2506279

[176] Shor P W. in *Proceedings of 37th Conference on Foundations of Computer Science.* 56.

[177] Wilczek F 2009 *Nat. Phys.* **5** 614

[178] He Q L, Pan L, Stern A L, Burks E C, Che X, Yin G, Wang J, Lian B, Zhou Q, Choi E S, Murata K, Kou X, Chen Z, Nie T, Shao Q, Fan Y, Zhang S-C, Liu K, Xia J and Wang K L 2017 *Science* **357** 294

[179] Young S M, Zaheer S, Teo J C Y, Kane C L, Mele E J and Rappe A M 2012 *Phys. Rev. Lett.* **108** 140405

[180] Fei Z, Palomaki T, Wu S, Zhao W, Cai X, Sun B, Nguyen P, Finney J, Xu X and Cobden D H 2017 *Nat. Phys.* **13** 677

[181] Tang S, Zhang C, Wong D, Pedramrazi Z, Tsai H-Z, Jia C, Moritz B, Claassen M, Ryu H, Kahn S, Jiang J, Yan H, Hashimoto M, Lu D, Moore R G, Hwang C-C, Hwang C, Hussain Z, Chen Y, Ugeda M M, Liu Z, Xie X, Devereaux T P, Crommie M F, Mo S-K and Shen Z-X 2017 *Nat. Phys.* **13** 683

[182] Tian W, Yu W, Liu X, Wang Y and Shi J. A Review of the Characteristics, Synthesis, and Thermodynamics of Type-II Weyl Semimetal WTe2. *Materials* **11**, 1185 (2018).

[183] Armitage N P, Mele E J and Vishwanath A 2018 *Rev. Mod. Phys.* **90** 015001

[184] Jia L-G, Liu M, Chen Y-Y, Zhang Y and Wang Y-L 2022 *Acta Physica Sinica* **71** 127308

[185] Wang M-X, Liu C, Xu J-P, Yang F, Miao L, Yao M-Y, Gao C L, Shen C, Ma X, Chen X, Xu Z-A, Liu Y, Zhang S-C, Qian D, Jia J-F and Xue Q-K 2012 *Science* **336** 52

[186] Huang C, Narayan A, Zhang E, Liu Y, Yan X, Wang J, Zhang C, Wang W, Zhou T, Yi C, Liu S, Ling J, Zhang H, Liu R, Sankar R, Chou F, Wang Y, Shi Y, Law K T, Sanvito S, Zhou P, Han Z and Xiu F 2018 *ACS Nano* **12** 7185

[187] Huang C, Narayan A, Zhang E, Xie X, Ai L, Liu S, Yi C, Shi Y, Sanvito S and Xiu F 2020 *Natl. Sci. Rev.* **7** 1468

[188] Lüpke F, Waters D, de la Barrera S C, Widom M, Mandrus D G, Yan J, Feenstra R M and Hunt B M 2020 *Nat. Phys.* **16** 526

[189] Bi X, Zhang Y, Ao L, Li H, Huang J, Qin F and Yuan H 2025 *Adv. Funct. Mater.* **35** 2415988

[190] Li C, de Boer J C, de Ronde B, Ramankutty S V, van Heumen E, Huang Y, de Visser A, Golubov A A, Golden M S and Brinkman A 2018 *Nat. Mater.* **17** 875

[191] Le T, Zhang R, Li C, Jiang R, Sheng H, Tu L, Cao X, Lyu Z, Shen J, Liu G, Liu F, Wang Z, Lu L and Qu F 2024 *Nat. Commun.* **15** 2785

[192] Yasuda K, Yasuda H, Liang T, Yoshimi R, Tsukazaki A, Takahashi K S, Nagaosa N, Kawasaki M and Tokura Y 2019 *Nat. Commun.* **10** 2734

[193] Masuko M, Kawamura M, Yoshimi R, Hirayama M, Ikeda Y, Watanabe R, He J J, Maryenko D, Tsukazaki A, Takahashi K S, Kawasaki M, Nagaosa N and Tokura Y 2022 *npj Quantum Mater.* **7** 104

[194] Zhang H, Liu C-X, Qi X-L, Dai X, Fang Z and Zhang S-C 2009 *Nat. Phys.* **5** 438

[195] Yang F, Ding Y, Qu F, Shen J, Chen J, Wei Z, Ji Z, Liu G, Fan J, Yang C, Xiang T and Lu L 2012 *Phys. Rev. B* **85** 104508

[196] Nadj-Perge S, Drozdov I K, Li J, Chen H, Jeon S, Seo J, MacDonald A H, Bernevig B A and Yazdani A 2014 *Science* **346** 602

[197] Xu J-P, Wang M-X, Liu Z L, Ge J-F, Yang X, Liu C, Xu Z A, Guan D, Gao C L, Qian D, Liu Y, Wang Q-H, Zhang F-C, Xue Q-K and Jia J-F 2015 *Phys. Rev. Lett.* **114** 017001

[198] Sun H-H, Zhang K-W, Hu L-H, Li C, Wang G-Y, Ma H-Y, Xu Z-A, Gao C-L, Guan D-D, Li Y-Y, Liu C, Qian D, Zhou Y, Fu L, Li S-C, Zhang F-C and Jia J-F 2016 *Phys. Rev. Lett.* **116** 257003

[199] Trang C X, Shimamura N, Nakayama K, Souma S, Sugawara K, Watanabe I, Yamauchi K, Oguchi T, Segawa K, Takahashi T, Ando Y and Sato T 2020 *Nat. Commun.* **11** 159

[200] Wan X, Turner A M, Vishwanath A and Savrasov S Y 2011 *Phys. Rev. B* **83** 205101

[201] Burkov A A and Balents L 2011 *Phys. Rev. Lett.* **107** 127205

[202] Lv B Q, Qian T and Ding H 2021 *Rev. Mod. Phys.* **93** 025002

[203] Wang Y, Liu E, Liu H, Pan Y, Zhang L, Zeng J, Fu Y, Wang M, Xu K, Huang Z, Wang Z, Lu H-Z, Xing D, Wang B, Wan X and Miao F 2016 *Nat. Commun.* **7** 13142

[204] Wang Y, Wang L, Liu X, Wu H, Wang P, Yan D, Cheng B, Shi Y, Watanabe K, Taniguchi T, Liang S-J and Miao F 2019 *Nano Lett.* **19** 3969

[205] Wang L, Xiong J, Cheng B, Dai Y, Wang F, Pan C, Cao T, Liu X, Wang P, Chen M, Yan S, Liu Z, Xiao J, Xu X, Wang Z, Shi Y, Cheong S-W, Zhang H, Liang S-J and Miao F 2022 *Sci. Adv.* **8** eabq6833

[206] Jindal A, Saha A, Li Z, Taniguchi T, Watanabe K, Hone J C, Birol T, Fernandes R M, Dean C R, Pasupathy A N and Rhodes D A 2023 *Nature* **613** 48

[207] Dai Y, Xiong J, Ge Y, Cheng B, Wang L, Wang P, Liu Z, Yan S, Zhang C, Xu X, Shi Y, Cheong S-W, Xiao C, Yang S A, Liang S-J and Miao F 2024 *Nat. Commun.* **15** 1129

[208] Li Q, He C, Wang Y, Liu E, Wang M, Wang Y, Zeng J, Ma Z, Cao T, Yi C, Wang N, Watanabe K, Taniguchi T, Shao L, Shi Y, Chen X, Liang S-J, Wang Q-H and Miao F 2018 *Nano Lett.* **18** 7962

[209] Lee J, Lee W, Kim G-Y, Choi Y-B, Park J, Jang S, Gu G, Choi S-Y, Cho G Y, Lee G-H and Lee H-J 2021 *Nano Lett.* **21** 10469

[210] Lee Y, Martini M, Confalone T, Shokri S, Saggau C N, Wolf D, Gu G, Watanabe K, Taniguchi T, Montemurro D, Vinokur V M, Nielsch K and Poccia N 2023 *Adv. Mater.* **35** 2209135

[211] Zhang D, Zhu Y-Y, Wang H and Xue Q-K 2023 *Acta Physica Sinica* **72** 237402

[212] Ghosh S, Patil V, Basu A, Kuldeep, Dutta A, Jangade D A, Kulkarni R, Thamizhavel A, Steiner J F, von Oppen F and Deshmukh M M 2024 *Nat. Mater.* **23** 612

[213] Wang H, Zhu Y, Bai Z, Lyu Z, Yang J, Zhao L, Zhou X J, Xue Q-K and Zhang D 2025 *Nat. Phys.* **22** 47

[214] Song Z, Qi J, Liebman O and Narang P 2024 *Phys. Rev. B* **110** 024401

[215] Zhu Y, Wang H, Wang Z, Hu S, Gu G, Zhu J, Zhang D and Xue Q-K 2023 *Phys. Rev. B* **108** 174508